# Globally Optimal Distributed Power Control for Nonconcave Utility Maximization

Li Ping Qian*, Ying Jun (Angela) Zhang*, and Mung Chiang+
*Department of Information Engineering, The Chinese University of Hong Kong
Shatin, New Territory, Hong Kong
+Department of Electrical Engineering, Princeton University, NJ 08544, USA
Email: *{lpqian6, yjzhang}@ie.cuhk.edu.hk +chiangm@princeton.edu

**Abstract**

Transmit power control in wireless networks has long been recognized as an effective mechanism to mitigate co-channel interference. Due to the highly non-convex nature, optimal power control is known to be difficult to achieve if a system utility is to be maximized. To date, there does not yet exist a distributed power control algorithm that maximizes any form of system utility, despite the importance of distributed implementation for the wireless infrastructureless networks such as ad hoc and sensor networks. This paper fills this gap by developing a Gibbs Sampling based Asynchronous distributed power control algorithm (referred to as GLAD). The proposed algorithm quickly converges to the global optimal solution regardless of the concavity, continuity, differentiability and monotonicity of the utility function. Same as other existing distributed power control algorithms, GLAD requires extensive message passing among all users in the network, which leads to high signaling overhead and high processing complexity. To address this issue, this paper further proposes a variant of the GLAD algorithm, referred to as I-GLAD, where the prefix "I" stands for infrequent message passing. The convergence of I-GLAD can be proved regardless of the reduction in the message passing rate. To further reduce the processing complexity at each transmitter, we develop an enhanced version of I-GLAD, referred to as NI-GLAD, where only the control messages from the neighboring links are processed. Our simulation results show that I-GLAD approximately converges to the global optimal solution regardless of the type of the system utility function. Meanwhile, the optimality of the solution obtained by NI-GLAD depends on the selection of the neighborhood size.

**Index Terms**

Power control, System utility maximization, Nonconvex global optimization.

This research is supported, in part, by the General Research Fund (Project number 419509) established under the University Grant Council of Hong Kong, and the National Natural Science Foundation of China (Project number: 61101132).

# I. Introduction

Due to the broadcast nature of wireless medium, simultaneous transmissions on nearby links cause severe co-channel interference to each other, thus adversely affecting the performance of the system. With the increasing density of wireless devices, interference mitigation has become a critical task to alleviate the adverse effect of co-channel interference. An important interference mitigation technique is to control the transmission power of links. One commonly pursued target of transmit-power control is to maximize a system wide efficiency metric, such as a system utility function based on throughput and delay [1]–[11].

Optimal power control is known to be difficult to achieve, mainly because the optimization problem is in general non-convex due to the complicated coupling among the signal-to-interference-and-noise ratios (SINRs) of different links [10]. A majority of efforts in tackling the problem is to convexify it through transformation, reparameterization, relaxation and approximation, oftentimes compromising the optimality of the solution [2], [3]. The first and probably the only global optimal power control scheme, referred to as MAPEL, that works for both concave and non-concave utility functions in all SINR regions was proposed in our recent work [11], [12]. In contrast to other previous work that tries to tackle the non-convexity issue head on, MAPEL bypasses non-convexity by exploiting the monotonic nature of the problem. The global optimal solution thus obtained provides important benchmark and guidelines for the evaluation and design of practical heuristics targeting the same problem.

In wireless infrastructureless networks such as ad hoc and sensor networks, power control is further complicated by the lack of centralized infrastructure, which necessitates the use of distributed approaches. Distributed power control is also often preferred in infrastructure networks, since coordination across base stations in multi-cellular networks is impractical. MAPEL, however, does not serve this purpose, as the monotonic optimization framework it adopts requires centralized coordination. Existing distributed power control algorithms are likely to converge to suboptimal solutions except for few special cases. For example, the algorithms in [8], [9] obtain the unique global optimal solution only when the utility function is strictly increasing, twice differentiable, and strictly log-concave in the feasible SINR region. For the many utility functions that do not satisfy these properties, including simple ones such as system throughput, the algorithms only converge to a KKT point that is not necessarily optimal. Another thread of work is to formulate distributed power control as a noncooperative pricing game [4]–[6]. Noticeably, various assumptions are imposed on the utility functions to ensure the existence of a unique Nash equilibrium. For example, utility functions are assumed to be logarithmic in SINR [5], differentiable and quasi-concave in transmission power [4], and increasing and concave in SINR [6]. However, due to the externalities of the power control problem[1], pricing-game based power control mechanisms generally fail to obtain the global optimal solution except for some special cases. An externalities-based decentralized algorithm is proposed in [7]. Therein, global optimality is obtained only when the utility function is continuous and strictly concave in transmission power, which is not the case in for most practical utility functions. To date, there does not yet exist a distributed power control algorithm that obtains the global optimal solution for general utility functions.

In this paper, we propose a distributed power control scheme that obtains the global optimal solution for any forms of utility functions. The algorithm, referred to as GLAD (Gibbs sampLing based Asynchronous Distributed) algorithm, is based on the idea of Gibbs Sampling, a reinforcement learning approach. Recently, Gibbs sampling has been applied to solve various optimization problems in wireless communication systems, such as minimizing end-to-end delays in mobile ad hoc networks [13], analyzing the optimality of the solution [14], and minimizing the sum of the inverse SINRs in cellular networks [15]. However, due to the lack of rigorous analysis, it is not clear how good the convergence rate and the solution quality are when Gibbs Sampling is applied to wireless networks. Moreover, all these existing algorithms were proposed for specific utility functions. In contrast with existing work [13]–[15], the GLAD algorithm proposed in this paper obtains *the optimal solution for general utility functions*. More importantly, this paper rigorously analyzes the performance of the GLAD algorithm, including the effect of a "temperature" parameter on the convergence rate and the solution quality. Through such analysis, we have better ideas on how to set the temperature parameter so that the algorithm operates in a desirable regime. In particular, our algorithm is distinguished from existing ones by the following features.

- First, the GLAD algorithm is guaranteed to converge to the optimal power control solution with any type of utility function regardless of the concavity, continuity, differentiability, monotonicity, and whether it is additive across links. GLAD is more widely applicable than its centralized counterpart, MAPEL, as the MAPEL algorithm requires the system utility to be monotonic in SINR.
- Second, GLAD exhibits faster convergence, compared with the MAPEL algorithm. In particular, our analysis shows that the algorithm converges linearly to the optimal solution.
- Third, GLAD allows asynchronous power update at each user. This is a crucial feature in wireless infrastructureless networks due to the lack of a central clock.

[1]A resource allocation problem is said to have externalities if the resources allocated to each user directly affect the utility of every other user.

- Last, for practical implementation of the algorithm, we further study how to set a "temperature" parameter of the algorithm. In particular, we provide rigorous study on the effect of the temperature parameter on the optimality of the solution as well as the effect on the convergence rate.

Similar to other distributed power control algorithms, the GLAD algorithm requires extensive message passing among users, which leads to high signaling overhead that negates the potential gains achieved by the algorithm. Meanwhile, these messages need to be received and decoded by the transmitter of each link. Thus, the large amount of message passing among users inevitably leads to high processing complexity at each transmitter.

To address this issue, this paper further proposes two variants of the GLAD algorithm, referred to as I-GLAD (the prefix "I" stands for infrequent message passing) and NI-GLAD (the prefix "N" stands for neighboring message passing), respectively. In particular, I-GLAD greatly reduces the frequency at which control messages are broadcasted, and thus significantly reduces the signaling overhead and processing complexity. In addition, NI-GLAD limits the message passing within a pre-defined neighborhood, and thus the processing complexity at each transmitter would not grow indefinitely with the network size. As a result, the scalability of the algorithm is greatly improved. Our analysis and simulation show that I-GLAD still converges despite the reduction in the message passing. The optimality of NI-GLAD, on the other hand, depends on the selection of the neighborhood size, which is quite intuitive.

The rest of this paper is organized as follows. Section II introduces the system model and problem formulation. The GLAD algorithm is presented in Section III. In Section IV, we analyze the effect of the temperature parameter on the optimality of the solution. In Section V, the convergence rate of the GLAD algorithm is presented. We extend the GLAD algorithm to the continuous power allocation case in Section VI. In Section VII, the variants of GLAD (i.e., I-GLAD and NI-GLAD) are proposed, and meanwhile their convergence is analyzed. The performance of GLAD, I-GLAD and NI-GLAD is evaluated through numerical simulations in Section VIII. Some discussions on the effect of the temperature parameter are also presented. The paper is concluded in Section IX.

## II. SYSTEM FORMULATION

We consider a snapshot of wireless ad hoc network with a set of *distinct* links denoted by $\mathcal{M} = \{1, \cdots, M\}$. Each link consists of a transmitter node $T_i$ and a receiver node $R_i$. The channel gain between node $T_i$ and node $R_j$ is denoted by $G_{ij}$, which is determined by various factors such as path loss, shadowing and fading effects. We write the channel gains into a matrix form $\boldsymbol{G} = [G_{ij}]$.

Let $p_i$ denote the transmission power of link $i$ (i.e., from node $T_i$), with $P_i^{\max}$ being its maximum allowable value. For notational convenience, we write $\boldsymbol{p} = (p_i, \forall i \in \mathcal{M})$ and $\boldsymbol{P}^{\max} = (P_i^{\max}, \forall i \in \mathcal{M})$ as the transmission power vector and the maximum transmission power vector, respectively. Likewise, let the noise received at $R_i$ be $n_i$. Thus, the received SINR of link $i$ is

$$\gamma_i(\boldsymbol{p}) = \frac{G_{ii} p_i}{\sum_{j \neq i} G_{ji} p_j + n_i}. \tag{1}$$

We aim to find the optimal power allocation $\boldsymbol{p}^*$ that maximizes the overall system utility $U(\boldsymbol{\gamma}(\boldsymbol{p}))$, where $\boldsymbol{\gamma}(\boldsymbol{p})$ is the vector of $\gamma_i(\boldsymbol{p})$. Mathematically, the power control problem is formulated into the following form:

$$\begin{aligned} \text{UM}: \quad & \underset{\boldsymbol{p}}{\text{maximize}} \quad U(\boldsymbol{\gamma}(\boldsymbol{p})) \\ & \text{subject to} \quad 0 \le p_i \le P_i^{\max}, \forall i \in \mathcal{M}. \end{aligned} \tag{2}$$

Due to the complicated coupling of SINR across links, Problem (UM) is in general non-convex even if the objective function $U(\cdot)$ is concave in $\boldsymbol{\gamma}(\boldsymbol{p})$, let alone the cases with non-concave $U(\cdot)$'s. It is worth noting that almost all previous work (e.g., [2], [3], [8], [9]) assumes that the system utility function $U(\cdot)$ is additive across links. That is, $U(\boldsymbol{\gamma}(\boldsymbol{p})) = \sum_{i=1}^{M} U_i(\gamma_i(\boldsymbol{p}))$, where $U_i(\cdot)$ is the utility of link $i$. There are, however, cases where the social welfare cannot be expressed as the summation of individual satisfaction.

Unlike the previous work, we do not impose any assumptions on the function $U(\cdot)$ except for $U(\cdot)$ being non-negative. In particular, $U(\cdot)$ is allowed to be non-additive, discontinuous, non-differentiable, non-concave, and non-monotonic. Thus, we have full freedom to choose the utility function $U(\cdot)$ that accurately reflects users' satisfaction.

## III. The Discrete-GLAD Algorithm

In this section, we develop a novel distributed algorithm, GLAD, based on Gibbs Sampling to solve Problem (UM). For the convenience of readers, we first review some preliminaries on Gibbs Sampling in Subsection III-A before presenting the Discrete-GLAD algorithm in Subsection III-B. Here, by discrete we mean transmission power can only be selected from a discrete and finite set.

### *A. Mathematical Preliminaries Related to Gibbs Sampling*

Gibbs Sampling was originally introduced by Gibbs in 1902 to model physical interactions between molecules and particles. Among its modern engineering applications is image processing optimization that maximizes the posterior mode estimate [16]. In particular, Gibbs Sampling solves an optimization problem with the following form

$$h^* = \min_{\boldsymbol{x} \in \mathcal{X}} \quad H(\boldsymbol{x}), \tag{3}$$

where the variable $\boldsymbol{x}$ is a $N$-dim row vector with element $x_n, n = 1, \cdots, N$, the feasible domain $\mathcal{X} = \prod_{n=1}^{N} \mathcal{X}_n \subset \mathcal{R}^N$ (the $N$-dim real domain) is a compact set from the Cartesian product of the discrete sets $\mathcal{X}_n$ corresponding to $x_n$, and the objective function $H(\boldsymbol{x})$ can be of any form.

The key idea of Gibbs Sampling is that the value of each $x_n$ is updated iteratively and asynchronously according to the probability distribution $\boldsymbol{\Lambda}_n(\boldsymbol{x}_{-n}) = (\Lambda_n(x_n \mid \boldsymbol{x}_{-n}), \forall x_n \in \mathcal{X}_n)$ with

$$\Lambda_n(x_n \mid \boldsymbol{x}_{-n}) = \frac{\exp(-\beta H(x_n, \boldsymbol{x}_{-n}))}{\sum_{x_n' \in X_n} \exp(-\beta H(x_n', \boldsymbol{x}_{-n}))}, \tag{4}$$

where $\boldsymbol{x}_{-n} = (x_1, \cdots, x_{n-1}, x_{n+1}, \cdots, x_N)$. Furthermore, the probability distribution $\boldsymbol{\Lambda}_n(\boldsymbol{x}_{-n})$ by itself is also adjusted at each iteration according to the observations of $x_1, \cdots, x_{n-1}, x_{n+1}, \cdots, x_N$. Presumably, the value of $x_n$ that yields a smaller $H(\boldsymbol{x})$ is more likely to be picked.

According to (4), an $x_n$ that yields a better objective function value (i.e., a smaller $H(\cdot)$ here) will be picked with a higher probability. This is especially true when $\beta$ is large. We will later come back to the pros and cons of setting $\beta$ large.

*B. Discrete-GLAD*

Assuming that each $p_i$ can only take values from a discrete set $\mathcal{P}_i^D = \{0, \Delta P_i, 2\Delta P_i, \cdots, P_i^{\max}\}$, Gibbs Sampling can be straightforwardly applied to solve Problem (UM) through minimizing the inverse of Utility $U(\cdot)$. In what follows, we present the Discrete-GLAD algorithm and prove its convergence to the set of global optimal solutions.

In Discrete-GLAD, each link $i$ picks a sequence of time epochs $\{t_i^{(1)}, t_i^{(2)}, \cdots\}$, at which its transmission power is updated. In particular, at time epoch $t_i^{(k)}$, the transmission power is updated to $p_i(t_i^{(k)})$ according to the probability distribution $\boldsymbol{\Lambda}_i(\boldsymbol{p}_{-i}(t_i^{(k)}-)) = (\Lambda_i(p_i \mid \boldsymbol{p}_{-i}(t_i^{(k)}-)), \forall p_i \in \mathcal{P}_i^D)$, i.e.,

$$\Lambda_i(p_i \mid \boldsymbol{p}_{-i}(t_i^{(k)}-)) = \frac{\exp(\frac{-\beta}{U(\boldsymbol{\gamma}(p_i,\boldsymbol{p}_{-i}(t_i^{(k)}-)))})}{\sum_{p_i' \in \mathcal{P}_i^D} \exp(\frac{-\beta}{U(\boldsymbol{\gamma}(p_i',\boldsymbol{p}_{-i}(t_i^{(k)}-)))})}, \tag{5}$$

where $\boldsymbol{p}_{-i}(t_i^{(k)}-) = (p_1(t_i^{(k)}-),\cdots,p_{i-1}(t_i^{(k)}-),p_{i+1}(t_i^{(k)}-),\cdots,p_M(t_i^{(k)}-))$ is the transmission power of all other links except link $i$ before the time instant $t_i^{(k)}$. Recall that $\gamma_j(\boldsymbol{p})$ is the SINR of link $j$ under transmission power vector $\boldsymbol{p}$. Then, the vectors $\gamma_j(p_i,\boldsymbol{p}_{-i}(t_i^{(k)}-))$ the SINR under transmission power vector $(p_i,\boldsymbol{p}_{-i}(t_i^{(k)}-))$, and $\boldsymbol{\gamma}(p_i,\boldsymbol{p}_{-i}(t_i^{(k)}-))$ is the vector of $\gamma_j(p_i,\boldsymbol{p}_{-i}(t_i^{(k)}-))$.

To calculate $\Lambda_i(p_i \mid \boldsymbol{p}_{-i}(t_i^{(k)}-))$ in (5), the transmitter of link $i$ needs to know $\gamma_j(p_i,\boldsymbol{p}_{-i}(t_i^{(k)}-))$ for all $j$. In principle, it can always calculate $\gamma_j(p_i,\boldsymbol{p}_{-i}(t_i^{(k)}-))$ through the definition of SINR in (1). However, this would require the transmitter $T_i$ to know the entire channel gain matrix $\boldsymbol{G}$ as well as the transmission power of all links. Luckily, a close observation shows that $T_i$ can calculate $\gamma_j(p_i,\boldsymbol{p}_{-i}(t_i^{(k)}-))$ according to the following equation:

$$\gamma_j(p_i,\boldsymbol{p}_{-i}(t_i^{(k)}-)) = \begin{cases} \frac{\gamma_j(\boldsymbol{p}(t_i^{(k)}-))p_i}{p_i(t_i^{(k)}-)}, & j = i \\ \frac{s_j(t_i^{(k)}-)}{\frac{s_j(t_i^{(k)}-)}{\gamma_j(\boldsymbol{p}(t_i^{(k)}-))} + G_{ij}(p_i - p_i(t_i^{(k)}-))}, & j \neq i, \end{cases} \tag{6}$$

where $s_j(t_i^{(k)}-) = G_{jj}p_j(t_i^{(k)}-)$ is the received signal power of link $j$ just before the time instant $t_i^{(k)}$. To facilitate the calculation in (6), all we need is for the receiver of each link $j$ to keep monitoring its SINR $\gamma_j$ and received signal power $s_j$. Once it senses a change in either of the two values, a control packet is broadcasted to inform other links of the new values. Thus, at time $t_i^{(k)}$, transmitter $T_i$ knows the information about $\gamma_j(\boldsymbol{p}(t_i^{(k)}-))$ and $s_j(t_i^{(k)}-)$ for all $j$, and can calculate $\gamma_j(p_i,\boldsymbol{p}_{-i}(t_i^{(k)}-))$ according to (6). Here, only the channel gain of its own link, $G_{ij}$, needs to be known by $T_i$.

Having introduced the basic operations, we now present the Discrete-GLAD algorithm in Table I. The convergence of the algorithm will be proved in the next section. Note that the convergence of Discrete-GLAD would still be guaranteed even if the links do not broadcast the information about $\gamma_j$ and $s_j$ in time. This will be discussed later.

## IV. THE OPTIMALITY OF THE SOLUTION

In this section, we investigate the optimality of the solution achieved by Discrete-GLAD. The following Theorem 1 shows that as $\beta$ approaches infinity, the algorithm converges to the one that picks global optimal solutions to Problem (UM) with probability 1. For finite $\beta$, Theorems 2, 3, and 4 show that the optimality of the solution obtained at convergence improves with the increase of $\beta$.

**Theorem 1.** *Starting from an arbitrary initial power allocation, Discrete-GLAD corresponds to a Markov chain that converges to a stationary distribution* $\boldsymbol{\Omega}_\beta = (\Omega_\beta(\boldsymbol{p}), \forall \boldsymbol{p} \in \mathcal{P}^D)$, *i.e.,*

$$\Omega_\beta(\boldsymbol{p}) = \frac{\exp(\frac{-\beta}{U(\gamma(\boldsymbol{p}))})}{\sum_{\boldsymbol{p}' \in \mathcal{P}^D} \exp(\frac{-\beta}{U(\gamma(\boldsymbol{p}'))})}, \tag{7}$$

*where* $\mathcal{P}^D = \{\boldsymbol{p} \mid p_i \in \mathcal{P}_i^D, \forall i\}$ *is the set of all feasible transmission power vectors. When* $\beta \to \infty$, $\Omega_\beta(\boldsymbol{p})$ *becomes*

$$\lim_{\beta \to \infty} \Omega_\beta(\boldsymbol{p}) = \Omega_\infty(\boldsymbol{p}) = \begin{cases} \frac{1}{|\mathcal{P}^{D*}|} & \text{if } \boldsymbol{p} \in \mathcal{P}^{D*}, \\ 0 & \text{otherwise,} \end{cases} \tag{8}$$

where $\mathcal{P}^{D*}$ is *the set of global optimal solutions to Problem (*UM*) with discrete power allocation, and* $|\mathcal{P}^{D*}|$ *denotes the cardinality of* $\mathcal{P}^{D*}$.

The proof of Theorem 1 is based on the model of Discrete-GLAD as a Markov chain with transition matrix $\boldsymbol{\Pi}$. The details are relegated to Appendix A.

Theorem 1 shows that as $\beta$ approaches infinity, Discrete-GLAD converges to a stationary distribution that selects the global optimal power vector with probability 1. When there are more than one global optimal solutions, they are selected equally likely. When $\beta$ is finite, the stationary distribution $\boldsymbol{\Omega}_\beta$ is nonzero for $\boldsymbol{p} \notin \mathcal{P}^{D*}$, implying that the non-optimal power vectors can also be selected. In the following theorems, we show that the optimality of the solution improves when $\beta$ increases.

**Theorem 2.** *For any optimal transmission power vector* $\boldsymbol{p}^*$, *the probability of choosing the optimal solution at convergence,* $\Omega_\beta(\boldsymbol{p}^*)$, *monotonically increases with the increase of* $\beta$.

*Proof:* To prove Theorem 2, we calculate the derivative of (7)

$$\frac{\partial \Omega_\beta(\boldsymbol{p}^*)}{\partial \beta} = \frac{\partial}{\partial \beta} \frac{\exp(\frac{-\beta}{U(\gamma(\boldsymbol{p}^*))})}{\sum_{\boldsymbol{p}' \in \mathcal{P}^D} \exp(\frac{-\beta}{U(\gamma(\boldsymbol{p}'))})} = \Omega_\beta(\boldsymbol{p}^*)(\langle U^{-1} \rangle_\beta - \frac{1}{U(\gamma(\boldsymbol{p}^*))}), \tag{9}$$

where $\left\langle U^{-1}\right\rangle_{\beta}$ denotes the expectation of the inverse utility over $\boldsymbol{\Omega}_{\beta}$, i.e., $\left\langle U^{-1}\right\rangle_{\beta}=\sum_{\boldsymbol{p}\in\mathcal{P}^{D}}\frac{\Omega_{\beta}(\boldsymbol{p})}{U(\gamma(\boldsymbol{p}))}$. Eqn. (9) shows that the sign of $\frac{\partial\Omega_{\beta}(\boldsymbol{p}^{*})}{\partial\beta}$ is determined by the sign of $\left\langle U^{-1}\right\rangle_{\beta}-\frac{1}{U(\gamma(\boldsymbol{p}^{*}))}$ as $\Omega_{\beta}(\boldsymbol{p}^{*})>0$ for any $\boldsymbol{p}^{*}$. Since $\frac{1}{U(\gamma(\boldsymbol{p}^{*}))}<\left\langle U^{-1}\right\rangle_{\beta}$ for every optimal solution $\boldsymbol{p}^{*}$, we have $\frac{\partial\Omega_{\beta}(\boldsymbol{p}^{*})}{\partial\beta}$>0. This implies that $\Omega_{\beta}(\boldsymbol{p}^{*})$ monotonically increases with $\beta$. ■

When the algorithm converges, the obtained system utility $U(\gamma(\boldsymbol{p}))$ is a random variable governed by the stationary distribution $\beta$. When $\beta\rightarrow\infty$, such random variable becomes deterministic and equals $U(\gamma(\boldsymbol{p}^{*}))$. Theorems 3 and 4 investigate the effect of $\beta$ on the expectation and variance of $U(\gamma(\boldsymbol{p}))$.

**Theorem 3.** *At convergence, the expected value of the system utility, denoted by* $\bar{U}_{\beta}=\sum_{\boldsymbol{p}\in\mathcal{P}^{D}}\Omega_{\beta}(\boldsymbol{p})U(\gamma(\boldsymbol{p}))$ *monotonically increases with the increase of* $\beta$. *Especially,* $\bar{U}_{\infty}$ *is equal to* $U(\gamma(\boldsymbol{p}^{*}))$ when $\beta\rightarrow\infty$.

*Proof:* Taking the derivative of $\bar{U}_{\beta}$, we get

$$\begin{aligned}\frac{\partial\bar{U}_{\beta}}{\partial\beta}&=\sum_{\boldsymbol{p}\in\mathcal{P}^{D}}\Omega_{\beta}(\boldsymbol{p})U(\gamma(\boldsymbol{p}))\left(\left\langle U^{-1}\right\rangle_{\beta}-\frac{1}{U(\gamma(\boldsymbol{p}))}\right)\\&=\sum_{\boldsymbol{p}\in\mathcal{P}^{D}}\Omega_{\beta}(\boldsymbol{p})U(\gamma(\boldsymbol{p}))\sum_{\boldsymbol{p}'\in\mathcal{P}^{D}}\frac{\Omega_{\beta}(\boldsymbol{p}')}{U(\gamma(\boldsymbol{p}'))}-1\\&>\left(\sum_{\boldsymbol{p}\in\mathcal{P}^{D}}\left(\Omega_{\beta}(\boldsymbol{p})U(\gamma(\boldsymbol{p}))\right)^{\frac{1}{2}}\left(\frac{\Omega_{\beta}(\boldsymbol{p})}{U(\gamma(\boldsymbol{p}))}\right)^{\frac{1}{2}}\right)^{2}-1\\&=\left(\sum_{\boldsymbol{p}\in\mathcal{P}^{D}}\Omega_{\beta}(\boldsymbol{p})\right)^{2}-1=1-1=0,\end{aligned}\tag{10}$$

where the inequality is due to the Holder's inequality. This implies that the mean $\bar{U}_{\beta}$ monotonically increases with the increase of $\beta$. When $\beta\rightarrow\infty$, $\Omega_{\beta}(\boldsymbol{p})$ is equal to $\frac{1}{|\mathcal{P}^{D*}|}$ for $\boldsymbol{p}\in\mathcal{P}^{D*}$. Thus,

$\bar{U}_{\infty}=\sum_{\boldsymbol{p}\in\mathcal{P}^{D*}}\frac{1}{|\mathcal{P}^{D*}|}U(\gamma(\boldsymbol{p}))=U(\gamma(\boldsymbol{p}^{*}))$. ■

In practice, we can only set $\beta$ to be a finite value. Corollary 2 says that $\beta$ should be at least larger than a threshold, if certain expected system utility is to be obtained. The corollary is straightforward from Theorem 3.

**Corollary 1.** For any $\bar{U}_{0}<\bar{U}<U(\gamma(\boldsymbol{p}^{*}))$, there exists a $\beta(\bar{U})$ such that the expected value of the obtained system utility at convergence (i.e., $\bar{U}_\beta$) is larger than $\bar{U}$ for all $\beta \geq \beta(\bar{U})$. In particular, $\beta(\bar{U})$ is given by

$$\bar{U}_{\beta(\bar{U})} = \bar{U} . \tag{11}$$

**Theorem 4.** *The variance of the system utility at convergence, denoted by* $V_\beta(U) = \sum_{\boldsymbol{p}\in\mathcal{P}^D} (U(\boldsymbol{\gamma}(\boldsymbol{p})) - \bar{U}_\beta)^2 \Omega_\beta(\boldsymbol{p})$, *is upper bounded by a decreasing function of* $\beta$, denoted by $\tilde{V}_\beta(U)$ *satisfies*

$$\tilde{V}_\beta(U) = U^{*2}\left(1 - |\mathcal{P}^{D*}|\,\Omega_\beta(\boldsymbol{p}^*)\right)^2 + (U^* - U_*)^2\left(1 - |\mathcal{P}^{D*}|\,\Omega_\beta(\boldsymbol{p}^*)\right), \tag{12}$$

where $U_* = \min_{\boldsymbol{p}\in\mathcal{P}^D} U(\boldsymbol{\gamma}(\boldsymbol{p}))$. When $\beta \to \infty$, $V_\beta(U) = \tilde{V}_\beta(U) = 0$.

We provide details of the proof in Appendix B.

Corollary 3 specifies a lower bound on $\beta$ if the variance of the obtained system utility is to be smaller than a threshold. It is straightforward from Theorem 4.

**Corollary 2.** For any $\delta \geq 0$, there exists a $\beta(\delta)$ such that the variance of the obtained utility at convergence (i.e.,$V_\beta(U)$) is smaller than $\delta$ for all $\beta \geq \beta(\delta)$. In particular, $\beta(\delta)$ is given by

$$\begin{cases} \beta(\delta) = 0, \quad if\ \delta \geq U^{*2}(1+\frac{|\mathcal{P}^{D*}|}{|\mathcal{P}^D|})^2 + (U^* - U_*)^2(1+\frac{|\mathcal{P}^{D*}|}{|\mathcal{P}^D|}), \\ \Omega_{\beta(\delta)}(\boldsymbol{p}) = \dfrac{1+\frac{(U^*-U_*)^2}{2U^{*2}} - \sqrt{\frac{\delta}{U^{*2}} + \frac{(U^*-U_*)^4}{4U^{*4}}}}{|\mathcal{P}^{D*}|}, \forall \boldsymbol{p} \in \mathcal{P}^{D*}, \text{otherwise.} \end{cases} \tag{13}$$

In summary, Theorems 2, 3, and 4 show that the optimality of the solution obtained at convergence improves with the increase of $\beta$. Moreover, Corollaries 2 and 3 specify the smallest value of $\beta$ we can choose if certain requirements on the expected value and variance of system utility is to be satisfied. This provides a guideline on the selection of $\beta$.

## V. The Rate of Convergence

Theorems 1 establishes the convergence of the Discrete-GLAD algorithm, but not the rate of convergence. In this section, we focus on the analysis of the convergence rate. One way to measure the convergence rate is to observe the total variation distance, defined in Definition 1, between two probability distributions $\boldsymbol{\Omega}_\beta^{(k)} = (\Omega_\beta^{(k)}(\boldsymbol{p}), \forall \boldsymbol{p} \in \mathcal{P}^D)$ and $\boldsymbol{\Omega}_\beta$, where $\boldsymbol{\Omega}_\beta^{(k)}$ is the probability distribution on the feasible power set $\mathcal{P}^D$ after the $k$ th update, and $\boldsymbol{\Omega}_\beta$ is the stationary distribution on the feasible power set $\mathcal{P}^D$ at the convergence of Discrete-GLAD.

**Definition 1.** The total variation distance between two probability distributions $\boldsymbol{\Omega}_\beta^{(k)}$ and $\boldsymbol{\Omega}_\beta$ is

defined as

$$\| \boldsymbol{\Omega}_{\beta}^{(k)} - \boldsymbol{\Omega}_{\beta} \|_{\text{var}} = \frac{1}{2} \sum_{\boldsymbol{p} \in \mathcal{P}^D} | \boldsymbol{\Omega}_{\beta}^{(k)}(\boldsymbol{p}) - \boldsymbol{\Omega}_{\beta}(\boldsymbol{p}) | . \tag{14}$$

**Theorem 5.** *Discrete-GLAD converges linearly to the stationary distribution* $\boldsymbol{\Omega}_{\beta}$ *in total variation distance. That is,*

$$\| \boldsymbol{\Omega}_{\beta}^{(k)} - \boldsymbol{\Omega}_{\beta} \|_{\text{var}} \le c_{\beta} | \lambda_{2\beta} |^{k} , \tag{15}$$

where $c_{\beta}$ is *a positive constant with respect to matrix* $\boldsymbol{\Pi}$ , $\lambda_{2\beta}$ *is the second largest eigenvalue of matrix* $\boldsymbol{\Pi}$ *satisfying* $0 < | \lambda_{2\beta} | < 1$.

The proof can be found in Appendix C.

## VI. THE CONTINUOUS-GLAD ALGORITHM

In the sections above, we have assumed that $p_i$ takes values from a finite and discrete set. However, the original Problem (UM) allows $p_i$ to be any real number in $[0, P_i^{\max}]$. One straightforward way to extend Discrete-GLAD to the continuous power allocation case is to let $\Delta P_i$ be very small (close to 0). However, a direct consequence is that each link now needs an excessively large memory space to store the probability distribution vector $\boldsymbol{\Lambda}_i(\boldsymbol{p}_{-i}(t_i^{(k)}-))$. To solve this issue, we have the following observation when $\Delta P_i \to 0$.

$$\Lambda_i(p_i \mid \boldsymbol{p}_{-i}(t_{ik}-)) = \lim_{\Delta P_i \to 0} \frac{\exp(\frac{-\beta}{U(\gamma(p_i, \boldsymbol{p}_{-i}(t_{ik}-)))}) \Delta P_i}{\Delta P_i \sum_{p_i' \in \mathcal{P}_i^D} \exp(\frac{-\beta}{U(\gamma(p_i', \boldsymbol{p}_{-i}(t_{ik}-)))})} = \lim_{\Delta P_i \to 0} \frac{\exp(\frac{-\beta}{U(\gamma(p_i, \boldsymbol{p}_{-i}(t_{ik}-)))}) \Delta P_i}{\int_0^{P_i^{\max}} \exp(\frac{-\beta}{U(\gamma(p_i', \boldsymbol{p}_{-i}(t_{ik}-)))}) dp_i'} . \tag{16}$$

This can be transformed into a probability density function (pdf)

$$f_i(p_i \mid \boldsymbol{p}_{-i}(t_{ik}-)) = \lim_{\Delta P_i \to 0} \frac{\Lambda_i(p_i, \boldsymbol{p}_{-i}(t_{ik}-))}{\Delta P_i} = \frac{\exp(\frac{-\beta}{U(\gamma(p_i, \boldsymbol{p}_{-i}(t_{ik}-)))})}{\int_0^{P_i^{\max}} \exp(\frac{-\beta}{U(\gamma(p_i', \boldsymbol{p}_{-i}(t_{ik}-)))}) dp_i'}, \forall p_i \in \mathcal{P}_i^C \tag{17}$$

where $\mathcal{P}_i^C = \{ p_i \mid 0 \le p_i \le P_i^{\max} \}$. With this, we extend Discrete-GLAD to its continuous counterpart, referred to as Continuous-GLAD. Each link $i$ updates its feasible transmission power $p_i \in \mathcal{P}_i^C$ at each time instant $t_i^{(k)}$ with the pdf (17). Continuous-GLAD algorithm is the same as Discrete-GLAD except for Step 7, which is modified as shown the Continuous-GLAD part of Table II.

The theorems and corollaries in Sections IV and V can be trivially extended to Continuous-GLAD. In the following, we restate the related theorems but omit the proofs due to space limitation. First, the convergence of Continuous-GLAD is given in the following Theorem 6.

**Theorem 6.** *Starting from any initial power allocation, Continuous-GLAD corresponds to a Markov chain that converges to a stationary joint pdf*

$$f_\beta(\boldsymbol{p}) = \frac{\exp(\frac{-\beta}{U(\gamma(\boldsymbol{p}))})}{\int_0^{\boldsymbol{P}^{\max}} \exp(\frac{-\beta}{U(\gamma(\boldsymbol{p}'))}) d\boldsymbol{p}'}, \forall \boldsymbol{p} \in \mathcal{P}^C \tag{18}$$

*where* $\mathcal{P}^C = \{\boldsymbol{p} \mid 0 \le p_i \le P_i^{\max}, \forall i \in M\}$. *When* $\beta \to \infty$, $f_\beta(\boldsymbol{p})$ *becomes*

$$\lim_{\beta \to \infty} f_\beta(\boldsymbol{p}) = f_\infty(\boldsymbol{p}) = \frac{1}{|\mathcal{P}^{C*}|} \delta_{\boldsymbol{p}}(\mathcal{P}^{C*}), \tag{19}$$

*where the function* $\delta_p(\mathcal{P}^{C*})$ *satisfies*

$$\delta_{\boldsymbol{p}}(\mathcal{P}^{C*}) = \begin{cases} +\infty, & if\ \boldsymbol{p} \in \mathcal{P}^{C*} \\ 0, & otherwise \end{cases} \tag{20}$$

and $\int_0^{\boldsymbol{P}^{\max}} \delta_{\boldsymbol{p}}(\mathcal{P}^{C*}) d\boldsymbol{p} = |\mathcal{P}^{C*}|$. *Here,* $\mathcal{P}^{C*}$ *is the set of global optimal solutions of Problem (*UM) *and* $|\mathcal{P}^{C*}|$ *denotes its cardinality. In other words, Continuous-GLAD converges to a strategy that selects global optimal power allocation with probability 1.*

Theorems 7 and 8 investigate the effect of $\beta$ on the expectation and variance of the obtained utility for the Continuous-GLAD algorithm.

**Theorem 7.** *When the algorithm converges, the expected value of obtained utility, denoted by* $\bar{U}_\beta = \int_0^{\boldsymbol{P}^{\max}} U(\gamma(\boldsymbol{p})) f_\beta(\boldsymbol{p}) d\boldsymbol{p}$, *monotonically increases with the increase of* $\beta$. *Especially,* $\bar{U}_\infty = U(\gamma(\boldsymbol{p}^*))$ when $\beta \to \infty$.

**Theorem 8.** *The variance of the system utility at convergence, denoted by* $V_\beta(U) = \int_0^{\boldsymbol{P}^{\max}} (U(\gamma(\boldsymbol{p})) - \bar{U}_\beta)^2 f_\beta(\boldsymbol{p}) d\boldsymbol{p}$, *is upper bounded by a decreasing function of* $\beta$ *denoted by* $\tilde{V}_\beta(U)$. *In particular,* $\tilde{V}_\beta(U)$ *satisfies*

$$\tilde{V}_\beta(U) = U^{*2}\left(1 - \int_{\boldsymbol{p} \in \mathcal{P}^{C*}} f_\beta(\boldsymbol{p}) d\boldsymbol{p}\right)^2 + (U^* - U_*)^2 \left(1 - \int_{\boldsymbol{p} \in \mathcal{P}^{C*}} f_\beta(\boldsymbol{p}) d\boldsymbol{p}\right). \tag{21}$$

When $\beta \to \infty$, $V_\beta(U) = \tilde{V}_\beta(U) = 0$.

Let $f_\beta^{(k)}(\boldsymbol{p})$ denote the joint pdf on the feasible transmission power set $\mathcal{P}^C$ when the *k*th update is completed, and let $\boldsymbol{\Pi}' = \lim_{\Delta P_i \to 0} \boldsymbol{\Pi}$. the convergence rate of Continuous-GLAD is given in the following Theorem 9.

**Theorem 9.** *Continuous-GLAD converges linearly to the stationary joint pdf* $f_\beta(\boldsymbol{p})$ *in total*

*variation distance. That is,*

$$\int_0^{\boldsymbol{P}^{\max}} | f_\beta^{(k)}(\boldsymbol{p}) - f_\beta(\boldsymbol{p}) | d\boldsymbol{p} \le c_\beta^{'} | \lambda_{2\beta}^{'} |^k, \tag{22}$$

*where* $c_\beta^{'}$ *is a positive constant with respect to matrix* $\boldsymbol{\Pi}^{'}$ *, and* $\lambda_{2\beta}^{'}$ *is the second largest eigenvalue of matrix* $\boldsymbol{\Pi}^{'}$ *in magnitude satisfying* $0 < | \lambda_{2\beta}^{'} | < 1$.

## VII. VARIANTS OF GLAD: I-GLAD AND NI-GLAD

The GLAD algorithm has several advantages over existing algorithms, such as convergence rate, global optimality and robustness. However, it requires each link $i$ to broadcast a control packet including $\gamma_i$ and $s_i$ once it senses a change in either of the two values. Although $s_i$ changes only when link $i$ updates its transmission power, $i$ changes whenever there is a link in the network updating its transmission power, which is much more frequent. Such frequent message passing generally leads to high signaling overhead and high processing complexity, which may negate the performance gains achieved. To address this issue, we propose two variants of GLAD: I-GLAD and NI-GLAD in this section. Meanwhile, through our analysis, we further prove the convergence of I-GLAD and NI-GLAD[2].

### A. GLAD with Infrequent Message Passing (I-GLAD)

The operation of I-GLAD is the same as GLAD except that a link broadcasts a control packet *only when* it updates its own transmission power. In other words, no control packet is sent when a link senses a change in its received SINR due to other links updating transmission powers. In this way, each link $i$ would still have the updated information about the received power $s_i$ 's of other links, but the information about $\gamma_i$ 's may be outdated. Let $\hat{\gamma}_j(t_i^{(k)}-)$ denote the received SINR announced by link $j$ in the last control packet sent before time instant $t_i^{(k)}$. This outdated information will be used by link $i$ when choosing its transmission power. In particular, the estimation of SINRs in (6) now becomes

$$\tilde{\gamma}_j(p_i, \boldsymbol{p}_{-i}(t_i^{(k)}-), \hat{\boldsymbol{\gamma}}(t_i^{(k)}-)) = \begin{cases} \dfrac{\hat{\gamma}_j(\boldsymbol{p}(t_i^{(k)}-))\, p_i}{p_i(t_i^{(k)}-)}, & j = i \\ \dfrac{s_j(t_i^{(k)}-)}{\dfrac{s_j(t_i^{(k)}-)}{\hat{\gamma}_j(\boldsymbol{p}(t_i^{(k)}-))} + G_{ij}(p_i - p_i(t_i^{(k)}-))}, & j \ne i, \end{cases} \tag{23}$$

[2]Since the original Problem (**UM**) allows $p_i$ to be any real number in $[[0, P_i^{\max}]$, we conduct the variants of GLAD in the case of continuous power allocation.

where $\hat{\boldsymbol{\gamma}}(t_i^{(k)}-)$ is the vector of $\hat{\gamma}_j(t_i^{(k)}-)$. Likewise, the pdf in (17) will now be calculated as

$$\tilde{f}_i(p_i \mid \boldsymbol{p}_{-i}(t_{ik}-), \hat{\boldsymbol{\gamma}}(t_i^{(k)}-)) = \frac{\exp(\frac{-\beta}{U(\tilde{\boldsymbol{\gamma}}(p_i, \boldsymbol{p}_{-i}(t_{ik}-), \hat{\boldsymbol{\gamma}}(t_i^{(k)}-)))})}{\int_0^{P_i^{\max}} \exp(\frac{-\beta}{U(\tilde{\boldsymbol{\gamma}}(p_i', \boldsymbol{p}_{-i}(t_{ik}-), \hat{\boldsymbol{\gamma}}(t_i^{(k)}-)))}) dp_i'}, \forall p_i \in \mathcal{P}_i^C, \quad (24)$$

where $\tilde{\boldsymbol{\gamma}}(p_i, \boldsymbol{p}_{-i}(t_i^{(k)}-), \hat{\boldsymbol{\gamma}}(t_i^{(k)}-))$ is the vector of $\tilde{\gamma}_j(p_i, \boldsymbol{p}_{-i}(t_i^{(k)}-), \hat{\boldsymbol{\gamma}}(t_i^{(k)}-))$. As a result, the I-GLAD algorithm differs from the original GLAD algorithm in the some steps (see the I-GLAD part of Table II).

**Remark 1.** In I-GLAD, the frequency at which control packets are broadcasted is reduced to $\frac{1}{M}$. This implies that the signaling overhead in I-GLAD is only $\frac{1}{M}$ of that in GLAD. Meanwhile, the number of control packets processed by each transmitter is also reduced to $\frac{1}{M}$.

In the following, we investigate the convergence of I-GLAD. Theorem 10 proves the convergence property of I-GLAD.

**Theorem 10.** *I-GLAD corresponds to a Markov chain that converges to a stationary distribution.*

The Proof is similar to the convergence proof of GLAD, and is given in Appendix D.

*B. I-GLAD with Neighboring Message Passing (NI-GLAD)*

The I-GLAD algorithm significantly reduces the volume of message passing. However, a transmitter $T_i$ still needs to process control packets from all $M$ links to calculate the vector $\tilde{\boldsymbol{\gamma}}(p_i, \boldsymbol{p}_{-i}(t_i^{(k)}-), \hat{\boldsymbol{\gamma}}(t_i^{(k)}-))$. That is, the processing complexity at each transmitter increases with the number of links, which in turn grows with the size (radius) of the network if we fix the density of links. In addition, the number of entries we need to calculate in each $\tilde{\boldsymbol{\gamma}}(p_i, \boldsymbol{p}_{-i}(t_i^{(k)}-), \hat{\boldsymbol{\gamma}}(t_i^{(k)}-))$, being $M$, also increases with the network size. Such increase in the computational and processing complexities limits the scalability of the algorithm.

Intuitively, links that are far away from each other hardly interfere. Thus, one way to deal with the aforementioned issue is to for each link to limit the attention to a small neighborhood. To do so, each link only processes control packet with a received SNR higher than a threshold $\bar{\gamma}$. That is, only control packets from close-by neighbors are processed.

Denote the set of neighboring links of link $i$ as $\mathcal{M}_i$. In particular, link $j$ belongs to $\mathcal{M}_i$ if control packet sent by link $j$ is received by link $i$ with the received SNR larger than $\bar{\gamma}$. By doing so, NI-GLAD remains the same as I-GLAD except that each link $i$ only has the knowledge of $\hat{\gamma}_j$ 's and $s_j$ 's for $j \in \mathcal{M}_i$. With this limited information, link $i$ evaluates $\tilde{\gamma}_j(p_i, \boldsymbol{p}_{-i}(t_i^{(k)}-), \hat{\boldsymbol{\gamma}}(t_i^{(k)}-))$'s for $j \in \mathcal{M}_i$ instead of for all $j \in \mathcal{M}$ at time $t_i^{(k)}$. Subsequently, it calculates the pdf as

$$\hat{f}_i(p_i \mid \boldsymbol{p}_{-i}(t_{ik}-), \hat{\boldsymbol{\gamma}}(t_i^{(k)}-)) = \frac{\exp(\frac{-\beta}{U(\tilde{\boldsymbol{\gamma}}_{\mathcal{M}_i}(p_i, \boldsymbol{p}_{-i}(t_{ik}-), \hat{\boldsymbol{\gamma}}(t_i^{(k)}-)))})}{\int_0^{P_i^{\max}} \exp(\frac{-\beta}{U(\tilde{\boldsymbol{\gamma}}_{\mathcal{M}_i}(p_i', \boldsymbol{p}_{-i}(t_{ik}-), \hat{\boldsymbol{\gamma}}(t_i^{(k)}-)))}) dp_i'}, \forall p_i \in \mathcal{P}_i^C \quad (25)$$

where $\tilde{\boldsymbol{\gamma}}_{\mathcal{M}_i}(\cdot)$ is the vector of $\tilde{\gamma}_j(\cdot)$ for all $j \in \mathcal{M}_i$. As a result, NI-GLAD differs from the I-GLAD algorithm in the some steps as shown in the NI-GLAD part of Table II.

**Remark 2.** In NI-GLAD, the computational and processing complexity at link $i$ is limited by the size of $\mathcal{M}_i$, which does not increase with $\mathcal{M}$ as long as the link density does not change.

In what follows, we investigate the convergence of NI-GLAD. The following Theorem 11 shows that given $\beta$, NI-GLAD converges to a strategy that picks feasible solutions to Problem (UM) with a unique stationary distribution.

**Theorem 11.** *NI-GLAD corresponds to a Markov chain that converges to a stationary distribution.*

*Proof:* The proof is similar to the proof of I-GLAD except for $\boldsymbol{\Pi}_i = [\Pi_i((p_1, \hat{\gamma}_1), (p_2, \hat{\gamma}_2)), \forall (p_1, \hat{\gamma}_1), (p_2, \hat{\gamma}_2) \in \mathcal{P}_\gamma)]$ with

$$\Pi_i((\boldsymbol{p}_1, \hat{\boldsymbol{\gamma}}_1), (\boldsymbol{p}_2, \hat{\boldsymbol{\gamma}}_2)) = \begin{cases} \hat{f}_i(p_{2,i} \mid \boldsymbol{p}_{1,-i}, \hat{\boldsymbol{\gamma}}_1), & \text{if } \boldsymbol{p}_{1,-i} = \boldsymbol{p}_{2,-i}, \hat{\boldsymbol{\gamma}}_{1,-i} = \hat{\boldsymbol{\gamma}}_{2,-i} \text{ and } \hat{\gamma}_{2,i} = \gamma(\boldsymbol{p}_2) \\ 0, & \text{otherwise,} \end{cases} \quad (26)$$

where $\mathcal{P}_\gamma$ is defined in Appendix D. Thus, the detailed proof is omitted in this report. ■

## VIII. Simulation Results

In this section, the performance of GLAD, I-GLAD and NI-GLAD is investigated through numerical simulations[3]. In particular, we assume the transmission power $p_i$ can take any continuous value from $[0, P_i^{\max}]$. Likewise, we assume the channel gain $G_{ij}$ is calculated based on two-ray ground reflection model, i.e., $G_{ij} = d_{ij}^{-4}$, where $d_{ij}$ denote the distance between node $T_i$ and node $R_j$.

[3]In this paper, the algorithms are implemented in MATLAB 7.0, and the time evaluation is based on a HP Compaq dx7300 desktop with 3.6GHz processors and 1Gb of RAM.

## A. Effect of $\beta$ on the Performance of GLAD

We first observe the effect of $\beta$ on the performance of GLAD. Consider an eight-link network with the channel gain matrix $\boldsymbol{G}$ as shown in Table III. For such a network, we run GLAD with different $\beta$. The attained system utility versus the number of iterations is plotted in Figs. 1 and 2. In particular, the proportional fairness utility $U(\boldsymbol{\gamma}(\boldsymbol{p}))=\prod_{i=1}^{M}\gamma_i(\boldsymbol{p})$ is considered in Fig. 1 and the total throughput utility $U(\boldsymbol{\gamma}(\boldsymbol{p}))=\sum_{i=1}^{M}\log_2(1+\gamma_i(\boldsymbol{p}))$ is considered in Fig. 2. Suppose that $P_i^{\max}=1.0\text{mW}$ and $n_i=0.1\mu\text{W}$ for each link $i$. For comparison purpose, the optimal system utility obtained by the MAPEL algorithm in [11] is also plotted.

From both figures, it can be seen that the average system utility when the algorithm converges improves with the increase of $\beta$. Moreover, the fluctuation (variance) in the system utility decreases as $\beta$ increases. This is because when $\beta$ is small, links tend to explore power allocations other than the optimal one from time to time, leading to larger fluctuation. The results are consistent with our analysis in Sections IV and VI.

To further illustrate the theorems, we plot in Fig. 3 the mean and the variance of the system utility achieved at convergence against $\beta$. Here, the system utility is the total throughput. Again, it can be seen that the mean in the achieved system utility increases with the increase of $\beta$. Specifically, the global optimal (maximum) system utility is achieved when $\beta\to\infty$. On the other hand, we can see that the variance (i.e., fluctuation) in the achieved system utility generally follows a decreasing trend and approaches 0 when $\beta$ is very large.

The above figures suggest that a larger $\beta$ is preferable when it comes to the optimality of the solution. However, it is not always the case when the convergence rate is of concern. Recall Fig. 2, and a close observation shows that a larger $\beta$ may lead to slower convergence. This is because the total throughput utility is non-concave in $\boldsymbol{p}$ and cannot be convexified through transformation. Thus, there exists more than one local optimal solution. Too large a $\beta$ may cause the GLAD algorithm to be too greedy, and consequently stuck in a local optimal solution for a long time before exploring other less greedy solutions. This phenomenon supports our discussions above. Note that such a phenomenon is not observed in Fig. 1. This is because the proportional fairness utility maximization problem can essentially be convexified through transformation [3]. Hence, there is only one optimal solution, which is the global optimal one, and the algorithm will not be stuck anywhere even if $\beta$ is large. Therefore, we can claim that the optimality of the solution and the convergence rate are both sensitive to the selection of $\beta$ for the non-concave system utility.

*B. Complexity Comparison with MAPEL*

In this subsection, we show that GLAD exhibits a faster convergence than its centralized counterpart MAPEL proposed in our earlier work [11]. First consider a six-link network, where the links are randomly placed in a 10m-by-10m area and the length of each link is uniformly distributed within the interval [1m, 2m]. Other system parameters are the same as those in the above figures. Fig. 4 shows that GLAD converges to the optimal solution in around 50 iterations (i.e., 10.34 seconds), while it takes MAPEL about 700 iterations (i.e., 210.8 seconds) to converge. As shown in our work [11], MAPEL can only be applied to the small-size networks (usually $M \leq 10$) due to its complexity. In contrast, GLAD can be applied to larger-size networks, as shown in Fig. 5. Here, we vary the total number of links $M$ from 11 to 20. The links are randomly located in a 30m-by-30m area. From the figure, we can see that the computational time of GLAD increases almost linearly with the increase of network size. As such, GLAD can efficiently handle a network size that is considered to be too large for the MAPEL algorithm.

Before leaving this subsection, we would like to emphasize that MAPEL can only handle system utilities that are monotonic in SINR. Meanwhile, no such constraint is imposed on GLAD.

*C. Performance Comparison between GLAD and I-GLAD*

In this subsection, we compare the performance of I-GLAD with GLAD under the eight-link network given in subsection VIII-B. The attained system utility versus the number of iterations is plotted in Figs. 6 and 7. In particular, the utilities adopted in Figs. 6 and 7 are the same as those in Figs. 1 and 2, respectively. Besides, all system parameters are the same as those in above figures. For comparison purpose, the optimal system utility obtained by the MAPEL algorithm in [11] is also plotted.

From both figures, it can be seen that the degradation of system utility as a result of infrequent message passing depends on the form of utility function. Specifically, Fig. 6 shows that for the proportional fairness utility, GLAD and I-GLAD achieves approximately the same performance. This may be because that the proportional fairness utility maximization problem can essentially be convexified through transformation, which makes the system utility insensitive the amount of message passing. However, such a phenomenon is not observed in Fig. 7, which shows that the fluctuation in the system utility increases with the reduction of message passing. This may be because the total throughput utility is non-concave in $\boldsymbol{p}$ and cannot be convexified through transformation, which makes the system utility sensitive to the reduction of message passing. We also find from Fig.7 that, although the fluctuation increases in I-GLAD, it still approximately converges to the global optimal solution.

### *D. Performance Comparison between GLAD and NI-GLAD*

Here, we compare the performance of NI-GLAD with GLAD through varying $\bar{\gamma}$. Here, the proportional fairness utility is considered in Fig. 8, and the total throughput utility is considered in Fig. 9. Other system parameters are the same as in the above figures. We consider a series of 15-link networks. The links are randomly located in a 50m-by-50m area and the length of each link is uniformly distributed within the interval [1m, 2m]. Each point is obtained by averaging over 100 different topologies of the same link density.

It is not surprising to see from Figs. 8 and 9 that the utility obtained by NI-GLAD depends on the size of the neighborhood, which decreases with the increase of $\bar{\gamma}$. Specifically, the obtained utility decreases with the increase of $\bar{\gamma}$. Interestingly, it can be seen from Figs. 8 and 9 that the degradation is a very small percentage even when $\bar{\gamma}$ is set to be 20dB. Such performance guarantee and low complexity make NI-GLAD preferable from practical implementation.

## IX. CONCLUSION

In this paper, we have proposed a distributed power control algorithm, GLAD, that efficiently converges to the set of global optimal solutions despite the non-convexity of power control problems. One key strength of the algorithm is that it applies to all forms of utility functions regardless of the concavity, continuity, differentiability, monotonicity, and whether it is additive across links. Moreover, it exhibits noticeably faster convergence than its centralized counterpart MAPEL. On the other hand, we have analyzed how a control parameter $\beta$ affects the algorithm performance. This helps to provide a theoretical guideline on how to set a value for $\beta$ that achieve the better solution quality.

To further improve the scalability but reduce the complexity of GLAD, we proposed two variants, namely I-GLAD and NI-GLAD. Both algorithms greatly enhance the practicality of the GLAD algorithm, while yielding a negligible performance degradation.

## APPENDIX A
## PROOF OF THEOREM 1

As stated in Algorithm 1, each link picks a sequence of time epochs to update its transmission power. In other words, a transition from the current transmission power vector $\boldsymbol{p}_1$ to another $\boldsymbol{p}_2$ occurs at one of the time epochs selected by any link. Here, we note several facts. First, the transition probability from $\boldsymbol{p}_1$ to $\boldsymbol{p}_2$ only depends on $\boldsymbol{p}_1$ but not the ones before $\boldsymbol{p}_1$. Second, with probability 1, no two links would update their transmission powers at the same time, because they schedule their power updates independently in continuous time. Last, with equal probability $\frac{1}{M}$, a transition is due to the power update of one link. As such, Discrete-GLAD can be modeled as a Markov chain, whose transition matrix is $\boldsymbol{\Pi} = [\Pi(\boldsymbol{p}_1, \boldsymbol{p}_2), \forall \boldsymbol{p}_1, \boldsymbol{p}_2 \in \mathcal{P}^D]$ satisfying

$$\boldsymbol{\Pi} = \frac{1}{M}\sum_{i=1}^{M} \boldsymbol{\Pi}_i, \tag{27}$$

where $\boldsymbol{\Pi}_i = [\Pi_i(\boldsymbol{p}_1, \boldsymbol{p}_2), \forall \boldsymbol{p}_1, \boldsymbol{p}_2 \in \mathcal{P}^D]$ with

$$\Pi_i(\boldsymbol{p}_1, \boldsymbol{p}_2) = \begin{cases} \Lambda_i(p_{2,i} \mid \boldsymbol{p}_{1,-i}), & \text{if } \boldsymbol{p}_{1,-i} = \boldsymbol{p}_{2,-i} \\ 0, & \text{otherwise.} \end{cases} \tag{28}$$

Here, $\boldsymbol{p}_{1,-i} = (p_{1,1}, \cdots, p_{1,i-1}, p_{1,i+1}, \cdots p_{1,M})$ is the transmission power vector of link $i$ 's opponents, and $\boldsymbol{p}_{2,-i}$ is defined likewise.

To prove the former part of Theorem 1, we just need to prove that the Markov chain specified in (27) has a stationary distribution. Note that there exists an index $k$ such that up to the $k$th update, all links have updated their transmission powers at least once. Thus, all entries in the matrix $\boldsymbol{\Pi}^{k'}$ are nonzero for $k' \geq k$. Thus, the Markov chain is irreducible, aperiodic and positive recurrent. This implies the existence of a stationary distribution. Meanwhile, the distribution $\boldsymbol{\Omega}_\beta$ given in (7) always satisfies $\boldsymbol{\Omega}_\beta \cdot \boldsymbol{\Pi} = \boldsymbol{\Omega}_\beta$. Consequently, $\boldsymbol{\Omega}_\beta$ is the stationary distribution of the Markov chain (i.e., the stationary distribution of Discrete-GLAD).

Now, we prove the latter part of Theorem 1. Let Let $U^* = \max_{p \in \mathcal{P}^D} U(\gamma(\boldsymbol{p})) = U(\gamma(\boldsymbol{p}^*))$ and $\Delta_{\frac{1}{U(\boldsymbol{p})}} = \frac{1}{U(\gamma(\boldsymbol{p}))} - \frac{1}{U^*}$. Note that $U(\gamma(\boldsymbol{p})) < U(\gamma(\boldsymbol{p}^*))$ for all $\boldsymbol{p} \notin \mathcal{P}^{D*}$. Then, it follows from (7) that

$$\Omega_\beta(\boldsymbol{p}) = \frac{\exp(-\beta \Delta_{\frac{1}{U(\boldsymbol{p})}})}{|\mathcal{P}^{D*}| + \sum\limits_{\boldsymbol{p}': U(\gamma(\boldsymbol{p}')) < U^*} \exp(-\beta \Delta_{\frac{1}{U(\boldsymbol{p})}})} \underset{\beta \to \infty}{\longrightarrow} \Omega_\infty(\boldsymbol{p}) = \begin{cases} \frac{1}{|\mathcal{P}^{D*}|} & \text{if } \boldsymbol{p} \in \mathcal{P}^{D*} \\ 0 & \text{otherwise} \end{cases} \tag{29}$$

Consequently, the latter part of Theorem 1 follows.

## APPENDIX B
## PROOF OF THEOREM 4

Note that

$$\bar{U}_\beta = \sum_{\boldsymbol{p}\in\mathcal{P}^{D*}} U(\gamma(\boldsymbol{p}))\Omega_\beta(\boldsymbol{p}) + \sum_{\boldsymbol{p}\notin\mathcal{P}^{D*}} U(\gamma(\boldsymbol{p}))\Omega_\beta(\boldsymbol{p}) \geq \sum_{\boldsymbol{p}\in\mathcal{P}^{D*}} U(\gamma(\boldsymbol{p}))\Omega_\beta(\boldsymbol{p}) \qquad (30)$$

due to the fact that $U(\cdot)$ is non-negative for any feasible power allocation. Since $\Omega_\beta(\boldsymbol{p})$'s and $U(\gamma(\boldsymbol{p}))$'s are equal to $\Omega_\beta(\boldsymbol{p}^*)$ and $U^*$ for all $\boldsymbol{p}\in\mathcal{P}^{D*}$, it follows from (30) that

$$\bar{U}_\beta \geq |\mathcal{P}^{D*}|\, U^*\Omega_\beta(\boldsymbol{p}^*). \qquad (31)$$

Then, we have

$$V_\beta(U) = |\mathcal{P}^{D*}|\left(U^* - \bar{U}_\beta\right)^2 \Omega_\beta(\boldsymbol{p}^*) + \sum_{\boldsymbol{p}\notin\mathcal{P}^{D*}} \left(U(\gamma(\boldsymbol{p})) - \bar{U}_\beta\right)^2 \Omega_\beta(\boldsymbol{p}) \qquad (32.1)$$

$$\leq |\mathcal{P}^{D*}|\, U^{*2}\left(1-|\mathcal{P}^{D*}|\,\Omega_\beta(\boldsymbol{p}^*)\right)^2 \Omega_\beta(\boldsymbol{p}^*) + (U^* - U_*)^2\left(1-|\mathcal{P}^{D*}|\,\Omega_\beta(\boldsymbol{p}^*)\right) \qquad (32.2)$$

$$\leq \underbrace{U^{*2}\left(1-|\mathcal{P}^{D*}|\,\Omega_\beta(\boldsymbol{p}^*)\right)^2 + (U^* - U_*)^2\left(1-|\mathcal{P}^{D*}|\,\Omega_\beta(\boldsymbol{p}^*)\right)}_{\tilde{V}_\beta(U)} \qquad (32.3)$$

Specifically, inequality (32.1) is due to (31) as well as the fact that the expected value $\bar{U}_\beta$ satisfies $U_* < \bar{U}_\beta \leq U^*$, and inequality (32.3) is due to the result of $\Omega_\beta(\boldsymbol{p}^*) \leq \frac{1}{|\mathcal{P}^{D*}|}$ from Corollary 1. Inqs. (32) show that $V_\beta(U)$ is upper bounded by $\tilde{V}_\beta(U)$. In fact, $\tilde{V}_\beta(U)$ is a decreasing function of $\beta$. This can be seen by noting that $\Omega_\beta(\boldsymbol{p}^*)$ monotonically increases with $\beta$ and $\tilde{V}_\beta(U)$ is a decreasing function of $\Omega_\beta(\boldsymbol{p}^*)$. Note that the upper bound $\tilde{V}_\beta(U)$ is tight when $\beta\to\infty$, as $\tilde{V}_\beta(U)$ itself equals 0 when $\beta\to\infty$. Thus, Theorem 4 follows.

## APPENDIX C
## PROOF OF THEOREM 5

Note that $\boldsymbol{\Omega}_\beta^{(k)} = \boldsymbol{V}\boldsymbol{\Pi}^k$ and $\boldsymbol{\Omega}_\beta = \boldsymbol{V}\boldsymbol{\Pi}^\infty$, where $\boldsymbol{V} = (V(\boldsymbol{p}), \forall \boldsymbol{p}\in\mathcal{P}^D)$ is the initial probability distribution on the feasible transmission power set $\mathcal{P}^D$ and $\boldsymbol{\Pi}^\infty$ is the limit of $\boldsymbol{\Pi}^k$ as $k\to\infty$. Note that, every entry $V(\boldsymbol{p})\geq 0$ and $\sum_{\boldsymbol{p}\in\mathcal{P}^D} V(\boldsymbol{p}) = 1$. Thus, $\|\boldsymbol{V}\|_{\text{var}} = \frac{1}{2}\sum_{\boldsymbol{p}\in\mathcal{P}^D} |V(\boldsymbol{p})| = \frac{1}{2}$. Correspondingly,

$$\|\boldsymbol{\Omega}_\beta^{(k)} - \boldsymbol{\Omega}_\beta\|_{\text{var}} = \|\boldsymbol{V}\boldsymbol{\Pi}^k - \boldsymbol{V}\boldsymbol{\Pi}^\infty\|_{\text{var}} \leq \|\boldsymbol{V}\|_{\text{var}} \cdot \|\boldsymbol{\Pi}^k - \boldsymbol{\Pi}^\infty\|_1 = \frac{1}{4}\|\boldsymbol{\Pi}^k - \boldsymbol{\Pi}^\infty\|_1. \qquad (33)$$

Therefore, to prove Theorem 5, we only need to investigate the property of $\|\boldsymbol{\Pi}^k - \boldsymbol{\Pi}^\infty\|_1$.

Let $\lambda_{j\beta}$ denote the *j*th eigenvalue of matrix $\boldsymbol{\Pi}$. With loss of generality, we assume that matrix $\boldsymbol{\Pi}$ has $l$ distinct eigenvalues, satisfying $|\lambda_{1\beta}|>|\lambda_{2\beta}|\geq\cdots\geq|\lambda_{l\beta}|$. Assuming that the algebraic multiplicity of eigenvalue $\lambda_{j\beta}$ is equal to $m_j$, we define a $m_j\times m_j$ matrix $\boldsymbol{J}_j$ as follows

$$\boldsymbol{J}_j=\begin{pmatrix}\lambda_{j\beta} & 1 & & 0\\ & \lambda_{j\beta} & \ddots & \\ & & \ddots & 1\\ 0 & & & \lambda_{j\beta}\end{pmatrix}=\lambda_{j\beta}\cdot\boldsymbol{I}_{m_j}+\underbrace{\begin{pmatrix}0 & 1 & & 0\\ & 0 & \ddots & \\ & & \ddots & 1\\ 0 & & & 0\end{pmatrix}}_{\boldsymbol{E}_j}, \tag{34}$$

where matrices $\boldsymbol{I}_{m_j}$ and $\boldsymbol{E}_j$ are an identity matrix and a nilpotent matrix[4], respectively. Likewise, we define a $|\mathcal{P}^D|\times m_j$ matrix $\boldsymbol{S}_j$ such that its *n*th column vector $\boldsymbol{s}_{j,n}$ satisfies $(\boldsymbol{\Pi}-\lambda_{j\beta}\boldsymbol{I}_{|\mathcal{P}^D|})^n\boldsymbol{s}_{j,n}=0$, where matrix $\boldsymbol{I}_{|\mathcal{P}^D|}$ is an identity matrix. According to the theory of Jordan matrix decomposition [17], the square matrix $\boldsymbol{\Pi}$ can then be decomposed into $\boldsymbol{\Pi}=\boldsymbol{SJS}^{-1}$, where $\boldsymbol{J}=\mathrm{diag}\{\boldsymbol{J}_1,\boldsymbol{J}_2,\cdots,\boldsymbol{J}_l\}$ and $\boldsymbol{S}=[\boldsymbol{S}_1,\boldsymbol{S}_2,\cdots,\boldsymbol{S}_l]$.

Perron-Frobenius theorem [17] says that, for a positive matrix, there exists a unique largest eigenvalue $\rho$ such that any other eigenvalue is strictly smaller than $\rho$ in absolute value. Meanwhile only eigenvalue $\rho$ is associated with a positive eigenvector. Note that $\boldsymbol{\Pi}$ is a positive matrix and has a positive eigenvector $\boldsymbol{\Omega}_\beta$ satisfying $\boldsymbol{\Omega}_\beta\cdot\boldsymbol{\Pi}=\boldsymbol{\Omega}_\beta$. Therefore, the unique largest eigenvalue of matrix $\boldsymbol{\Pi}$ is 1. That is, $\lambda_{1\beta}=1>|\lambda_{2\beta}|\geq\cdots\geq|\lambda_{l\beta}|$. Thus, $\boldsymbol{J}_1=[1]$ and $\boldsymbol{S}_1$ is a $|\mathcal{P}^D|\times 1$ matrix.

Taking the *k*th power of $\boldsymbol{\Pi}$, we have

$$\boldsymbol{\Pi}^k=\boldsymbol{S}\begin{pmatrix}1 & 0 & & 0\\ 0 & \boldsymbol{J}_2^k & & \\ & & \ddots & \\ 0 & & & \boldsymbol{J}_l^k\end{pmatrix}\boldsymbol{S}^{-1}\underset{k\to\infty}{\rightarrow}\boldsymbol{\Pi}^\infty=\boldsymbol{S}\underbrace{\begin{pmatrix}1 & 0 & & 0\\ 0 & 0 & & \\ & & \ddots & \\ 0 & & & 0\end{pmatrix}}_{z}\boldsymbol{S}^{-1}. \tag{35}$$

Since $|\lambda_{j\beta}|<1$ for $j\geq 2$, we get

$$\begin{aligned}\boldsymbol{J}_j^k&=[\lambda_{j\beta}\cdot\boldsymbol{I}_{m_j}+\boldsymbol{E}_j]^k=\lambda_{j\beta}^k\cdot\boldsymbol{I}_{m_j}+\frac{k\lambda_{j\beta}^k}{\lambda_{j\beta}}\cdot\boldsymbol{E}_j+\cdots+\frac{k!\lambda_{j\beta}^k}{(m_j-1)!(k-m_j+1)!\lambda_{j\beta}^{m_j-1}}\cdot\boldsymbol{E}_j^{m_j-1}\\ &\leq C_\beta\lambda_{j\beta}^k[\boldsymbol{I}_{m_j}+\boldsymbol{E}_j+\cdots+\boldsymbol{E}_j^{m_j-1}]\end{aligned} \tag{36}$$

for $j\geq 2$. Here, $C_\beta>0$. This implies that any $\boldsymbol{J}_j^k$ satisfies

$$\boldsymbol{J}_j^k \le C_\beta \lambda_{2\beta}^k [\boldsymbol{I}_{m_j} + \boldsymbol{E}_j + \cdots + \boldsymbol{E}_j^{m_j - 1}], \tag{37}$$

since $|\lambda_{2\beta}| \ge \cdots \ge |\lambda_{l\beta}|$. Therefore,

$$\begin{aligned} \| \boldsymbol{\Pi}^k - \boldsymbol{\Pi}^\infty \|_1 &\le \| \boldsymbol{S} \|_1 \cdot \| \boldsymbol{J}^k - \boldsymbol{Z} \|_1 \cdot \| \boldsymbol{S}^{-1} \|_1 \le \| \boldsymbol{I}_{|\mathcal{P}^D|} \|_1 \cdot \| \boldsymbol{J}^k - \boldsymbol{Z} \|_1 \\ &\le \underbrace{C_\beta |\mathcal{P}^D| \sum_{j=2}^{l} \frac{m_j(m_j+1)}{2}}_{c_\beta} |\lambda_{2\beta}|^k, \end{aligned} \tag{38}$$

where matrix $\boldsymbol{Z}$ is denoted in (35). Together with (33), it follows $\| \boldsymbol{\Omega}_\beta^{(k)} - \boldsymbol{\Omega}_\beta \|_{\text{var}} \le c_\beta |\lambda_{2\beta}|^k$, i.e. (15).

## APPENDIX D

## PROOF OF THEOREM 10

As a variant of GLAD, I-GLAD can also be modeled as a Markov chain with the continuous state space. Unlike GLAD, the transition probability for I-GLAD from $\boldsymbol{p}_1$ to $\boldsymbol{p}_2$ only depends on the state $(\boldsymbol{p}_1, \hat{\boldsymbol{\gamma}}_1)$ but not the ones before $(\boldsymbol{p}_1, \hat{\boldsymbol{\gamma}}_1)$. Here, $\hat{\boldsymbol{\gamma}}_1$ corresponds to the SINR vector known by each link at the selected time epoch. Each entry of the transition kernel for I-GLAD is thus just a conditional probability density of $(\boldsymbol{p}_2, \hat{\boldsymbol{\gamma}}_2)$ given $(\boldsymbol{p}_1, \hat{\boldsymbol{\gamma}}_1)$.

Next, we define the feasible domain of $(\boldsymbol{p}, \hat{\boldsymbol{\gamma}})$. Noticeably, the SINR vector $\hat{\boldsymbol{\gamma}}$ depends on not only the current transmission power vector $\boldsymbol{p}$ but also the outdated transmission power of each link. Assume the latest time of power updating for link $i$ is $t_i$. According to the order of latest updating time epochs $t_i$'s (i.e., $t_{i_1} < t_{i_2} < \cdots < t_{i_M}$), we construct a sequence $\{i_1, i_2, \cdots, i_M\}$, where $i_j$ represents the index of link. Specifically, the sequence $\{i_1, i_2, \cdots, i_M\}$ is a permutation of $\{1, 2, \cdots, M\}$. Suppose that the transmission power vector is $\boldsymbol{p}$ at time epoch $t_{i_M}$. As mentioned in I-GLAD, the updating order implies that the transmission power $p_{i_j}$ is picked following the transmission power $p_{i_{j-1}}$. Therefore, the transmission power of links $i_1, i_2, \cdots, i_j$ corresponds to $p_{i_1}, p_{i_2}, \cdots,$ and $p_{i_j}$ at time epoch $t_{i_j}$, respectively. Since $\hat{\gamma}_{i_j}$ is broadcasted at time epoch $t_{i_j}$, $\hat{\gamma}_{i_j}$ is calculated as

$$\hat{\gamma}_{i_j} = \gamma_{i_j}(p_{i_1}, \cdots, p_{i_j}, \tilde{p}_{i_{j+1}}, \cdots, \tilde{p}_{i_M}), \tag{39}$$

where $\tilde{p}_{i_{j+1}}, \cdots, \tilde{p}_{i_M}$ denote the transmission power of links from $i_{j+1}, \cdots, i_M$ at time epoch $t_{i_j}$. Furthermore, due to the fact that only one link updates its transmission power at the selected time epoch, only the transmission power of link $i_j$ is different at time epochs $t_{i_{j-1}}$ and $t_{i_j}$. Let $\hat{\boldsymbol{\gamma}}$ denote the vector $(\hat{\gamma}_{i_1}, \cdots, \hat{\gamma}_{i_M})$. Then, the domain of $(\boldsymbol{p}, \hat{\boldsymbol{\gamma}})$, denoted by $\mathcal{P}_\gamma$, can be formulated as $\mathcal{P}_\gamma = \{(\boldsymbol{p}, \hat{\boldsymbol{\gamma}}) \mid \boldsymbol{p} \in \mathcal{P}^D$, and $\hat{\boldsymbol{\gamma}}$ satisfies (39)$\}$. Now, we can express the transition kernel in $\boldsymbol{\Pi} = [\Pi((\boldsymbol{p}_1, \hat{\boldsymbol{\gamma}}_1), (\boldsymbol{p}_2, \hat{\boldsymbol{\gamma}}_2)), \forall (\boldsymbol{p}_1, \hat{\boldsymbol{\gamma}}_1), (\boldsymbol{p}_2, \hat{\boldsymbol{\gamma}}_2) \in \mathcal{P}_\gamma]$ satisfying

$$\boldsymbol{\Pi} = \frac{1}{M} \sum_{i=1}^{M} \boldsymbol{\Pi}_i , \qquad (40)$$

where $\boldsymbol{\Pi}_i = [\Pi_i((\boldsymbol{p}_1, \hat{\boldsymbol{\gamma}}_1), (\boldsymbol{p}_2, \hat{\boldsymbol{\gamma}}_2)), \forall (\boldsymbol{p}_1, \hat{\boldsymbol{\gamma}}_1), (\boldsymbol{p}_2, \hat{\boldsymbol{\gamma}}_2) \in \mathcal{P}_\gamma]$ with

$$\Pi_i((\boldsymbol{p}_1, \hat{\boldsymbol{\gamma}}_1), (\boldsymbol{p}_2, \hat{\boldsymbol{\gamma}}_2)) = \begin{cases} \tilde{f}_i(p_{2,i} \mid \boldsymbol{p}_{1,-i}, \hat{\boldsymbol{\gamma}}_1), & \text{if } \boldsymbol{p}_{1,-i} = \boldsymbol{p}_{2,-i},\ \hat{\boldsymbol{\gamma}}_{1,-i} = \hat{\boldsymbol{\gamma}}_{2,-i}, \text{ and } \hat{\gamma}_{2,i} = \gamma_i(\boldsymbol{p}) \\ 0, & \text{otherwise.} \end{cases} \qquad (41)$$

[4]The non-negative square matrix $\boldsymbol{E}_j$ is defined to be nilpotent, since there exists an integer $m_j > 0$ such that $\boldsymbol{E}^{m_j} = 0$.

To prove Theorem 10, we just need to show the Markov chain specified in (40) has a stationary distribution. Note that there exists an index $k$ such that up to the $k$th update, all links have updated their transmission powers at least once. Thus, all entries in the matrix $\boldsymbol{\Pi}^{k'}$ are nonzero for $k' \geq k$. Thus, the Markov chain is irreducible, aperiodic and positive recurrent. This implies the existence of a stationary distribution. Consequently, Theorem 10 follows.

TABLE I

ALGORITHM 1: THE DISCRETE-GLAD ALGORITHM

---

**The implementation at each transmitter node $T_i$:**

1: **Initialization:** pick a sequence of time epochs $\{t_i^{(1)}, t_i^{(2)}, \cdots\}$ in continuous time.
2: Choose some feasible power $p_i(t_i^{(1)}) \in \mathcal{P}_i^D$. Let $k = 1$.
3: **repeat**
4: Transmit the data packet with the power level $p_i(t_i^{(k)})$.
5: Keep sensing the control packets broadcasted by receivers, and then update the information of $\gamma_j$'s and $s_j$'s.
6: $k = k + 1$.
7: Update the feasible power $p_i(t_i^{(k)}) \in \mathcal{P}_i^D$ according to the probability distribution given in (5).
8: **until** Link $i$ decides to leave the network

**The implementation at each receiver node $R_i$:**

1: **repeat**
2: Keep measuring its received SINR and received power, and broadcast them in a control packet when a change in the SINR or power is sensed.
3: **until** Link $i$ leaves the network

---

TABLE II

ALGORITHMS: CONTINUOUS-GLAD, I-GLAD AND NI-GLAD

---

**Continuous-GLAD:**

*Step 7.* Choose a feasible power $p_i(t_i^{(k)}) \in \mathcal{P}_i^C$ according to the pdf given in (17).

---

**I-GLAD:**

**The implementation at each transmitter node $T_i$**

*Step 7:* Update the feasible power $p_i(t_i^{(k)}) \in \mathcal{P}_i^C$ according to the pdf given in (24).

**The implementation at each receiver node $R_i$**

*Step 2:* Keep measuring its received SINR and received power, and broadcast them in a control packet only when link $i$ updates its own transmission power.

---

**NI-GLAD:**

**The implementation at each transmitter node $T_i$**

*Step 7:* Update the feasible power $p_i(t_i^{(k)}) \in \mathcal{P}_i^C$ according to the pdf given in (25).

TABLE III

THE CHANNEL GAIN MATRIX $G$ USED IN SECTION VIII.A

$$G = \begin{bmatrix} 0.1116 & 0.0001 & 0.0040 & 0.0634 & 0.0004 & 0.0004 & 0.0012 & 0.0001 \\ 0.0001 & 0.4939 & 0.0004 & 0.0002 & 0.0411 & 0.0064 & 0.0046 & 0.0024 \\ 0.0004 & 0.0003 & 0.1586 & 0.0039 & 0.0015 & 0.0043 & 0.0006 & 0.0013 \\ 0.0185 & 0.0001 & 0.0159 & 0.7325 & 0.0006 & 0.0007 & 0.0013 & 0.0002 \\ 0.0001 & 0.0359 & 0.0011 & 0.0003 & 0.2913 & 0.1818 & 0.0024 & 0.0316 \\ 0.0001 & 0.0127 & 0.0010 & 0.0002 & 0.0321 & 0.1142 & 0.0010 & 0.4109 \\ 0.0002 & 0.0056 & 0.0007 & 0.0003 & 0.0206 & 0.0034 & 0.1887 & 0.0007 \\ 0.0001 & 0.0040 & 0.0003 & 0.0001 & 0.0021 & 0.0037 & 0.0003 & 0.1041 \end{bmatrix}$$

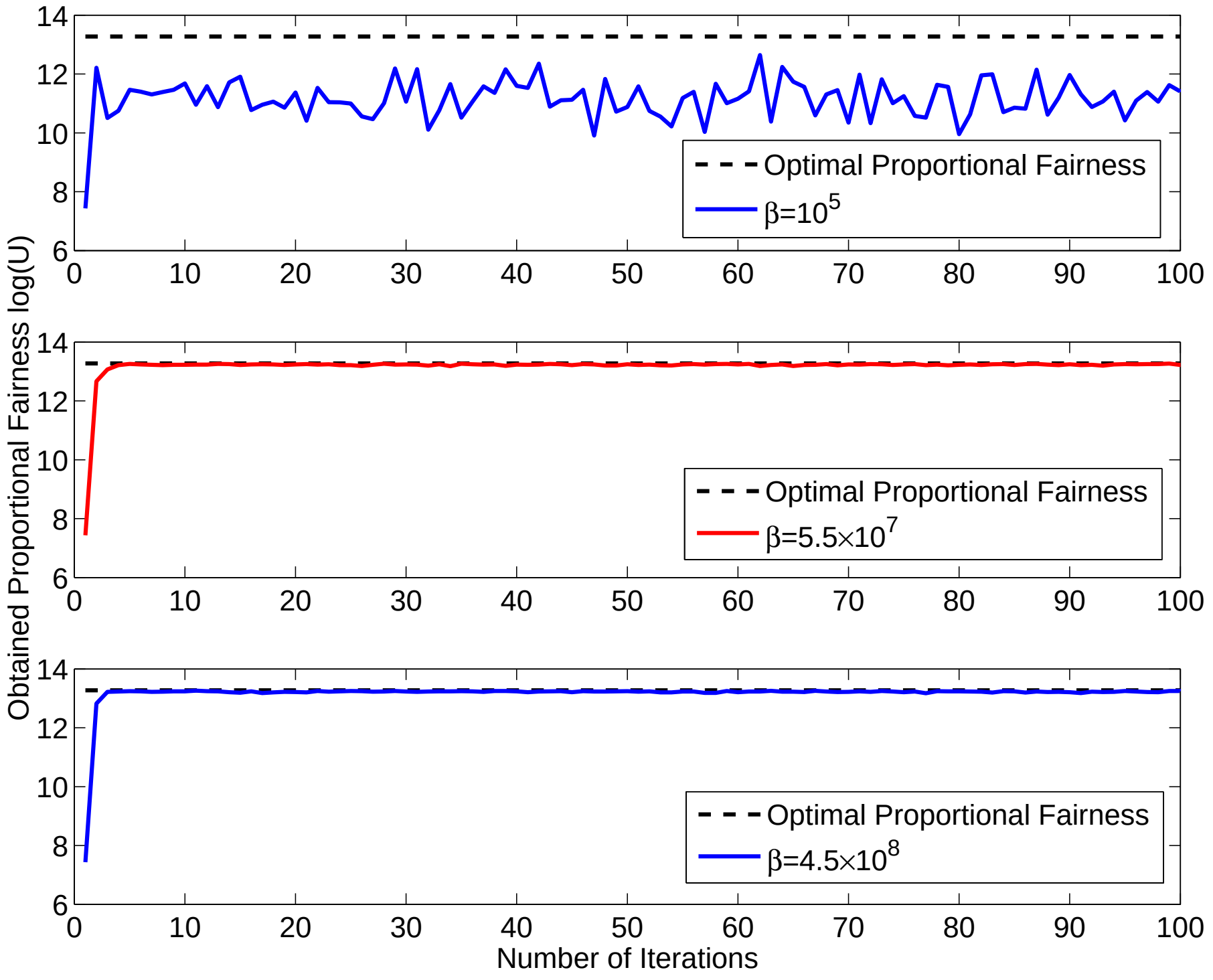

Fig. 1. Obtained Proportional Fairness and number of iterations for different $\beta$.

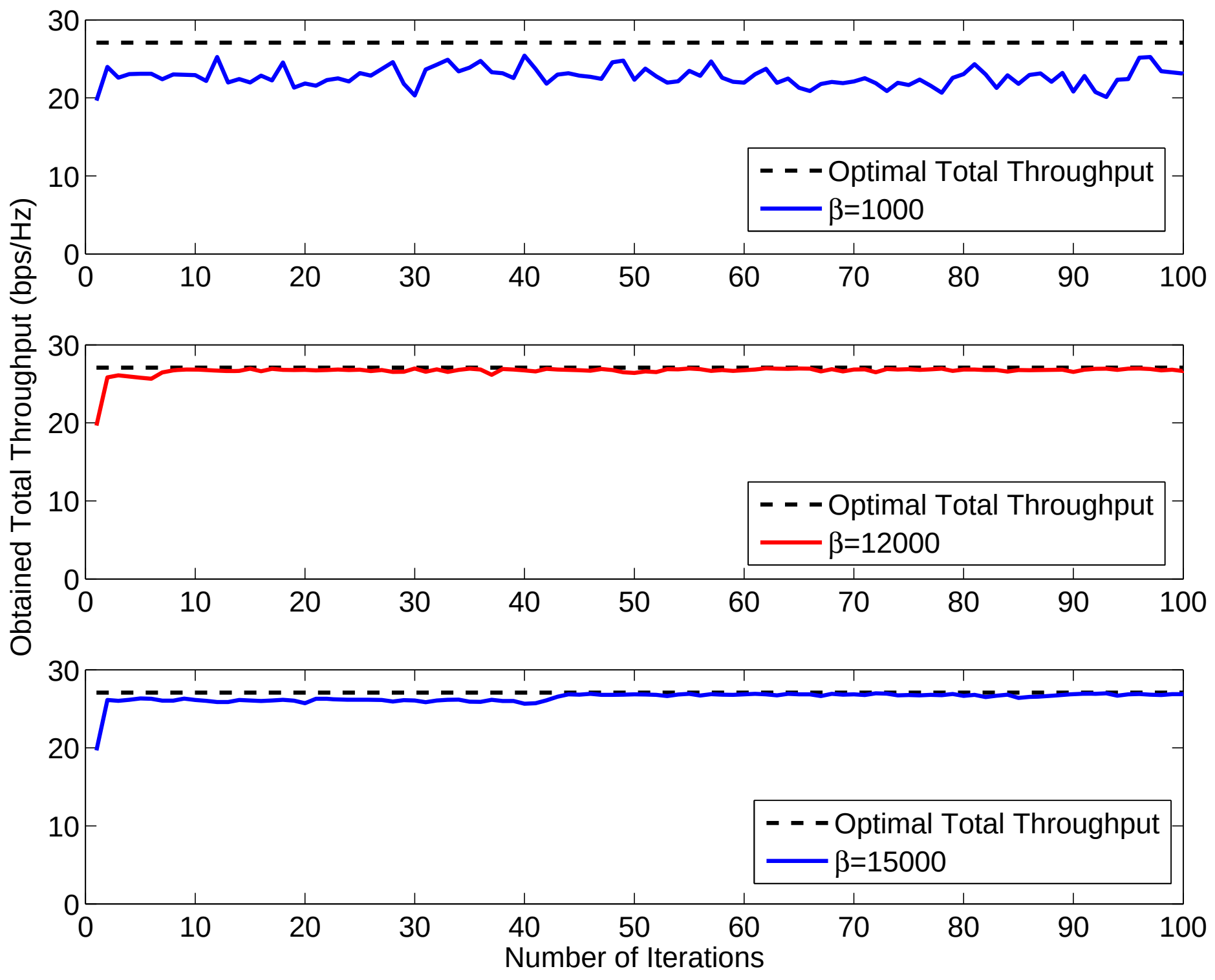

Fig. 2. Obtained Total Throughput and number of iterations for different $\beta$.

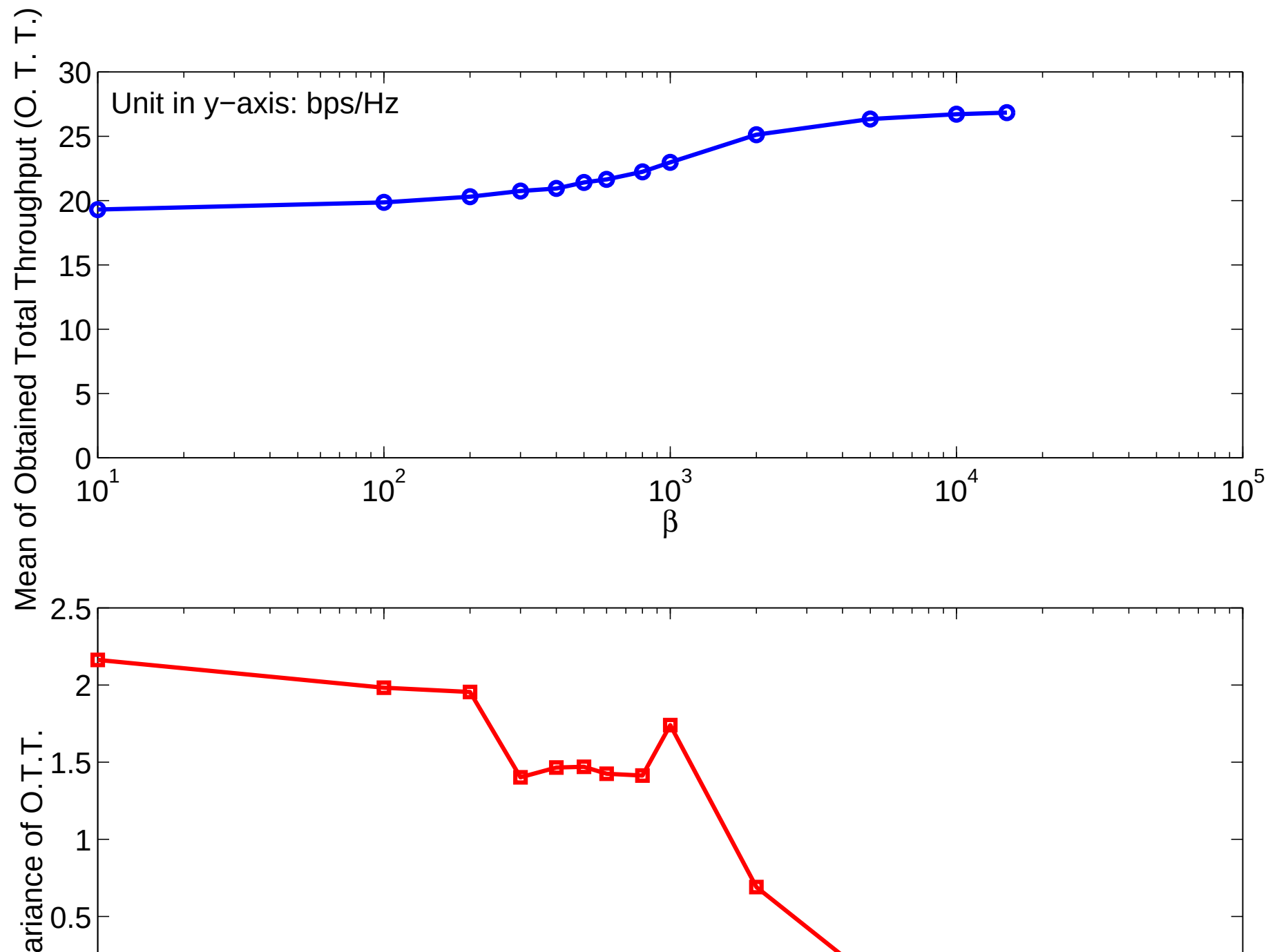

Fig. 3. The mean and the variance in the achieved system utility for different $\beta$.

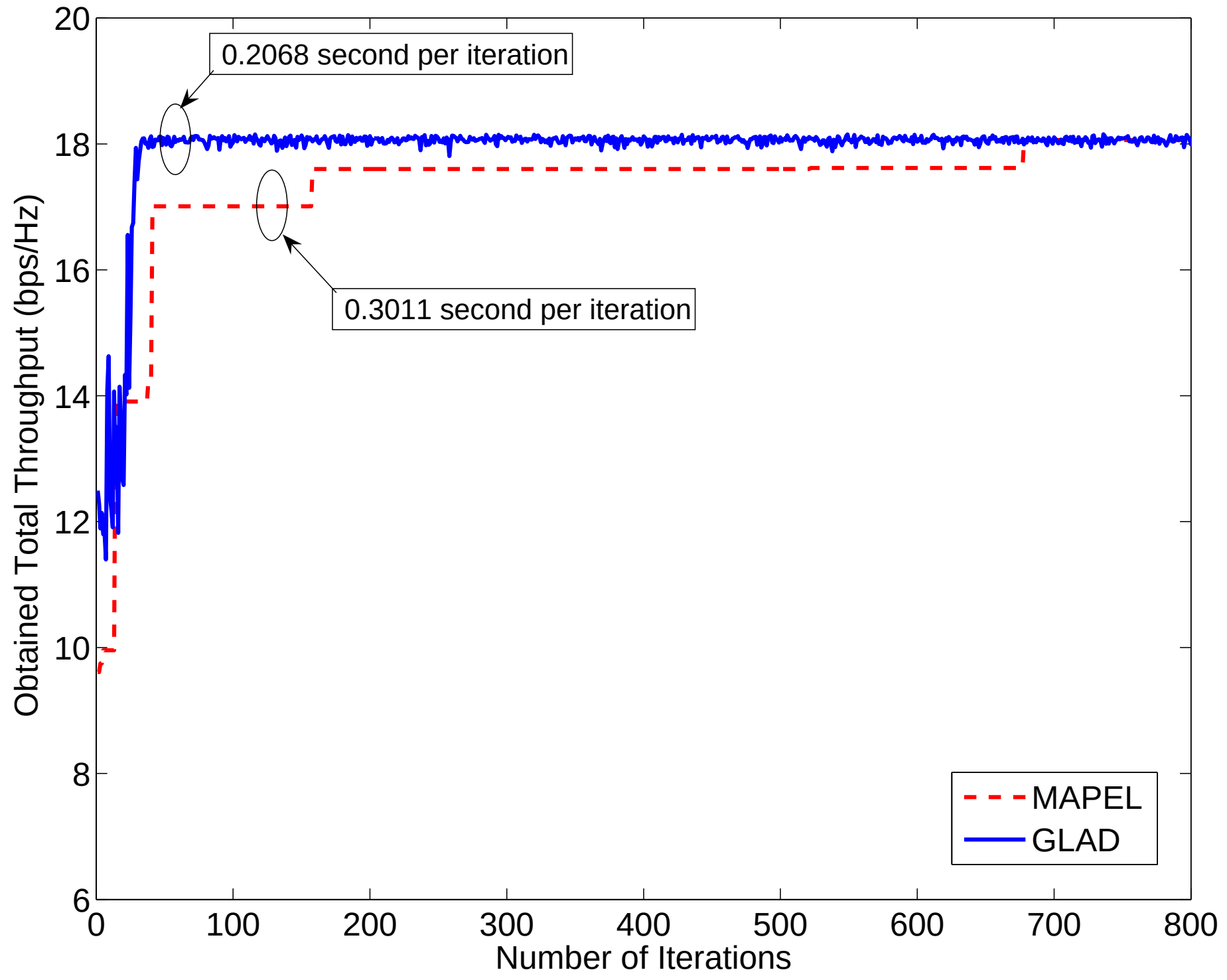

Fig. 4. The complexity comparison between GLAD and MAPEL.

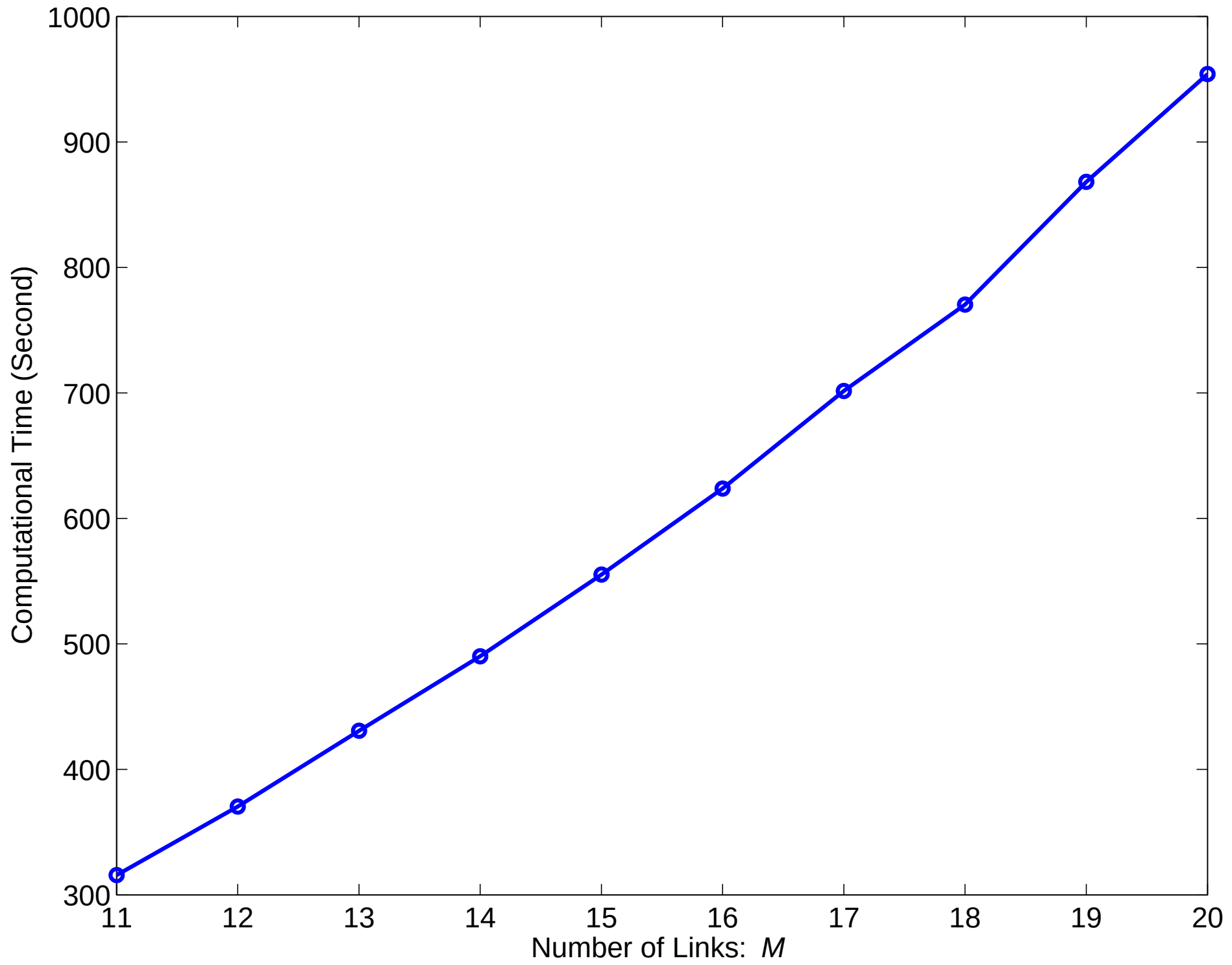

Fig. 5. The relationship between the complexity and the network size for GLAD.

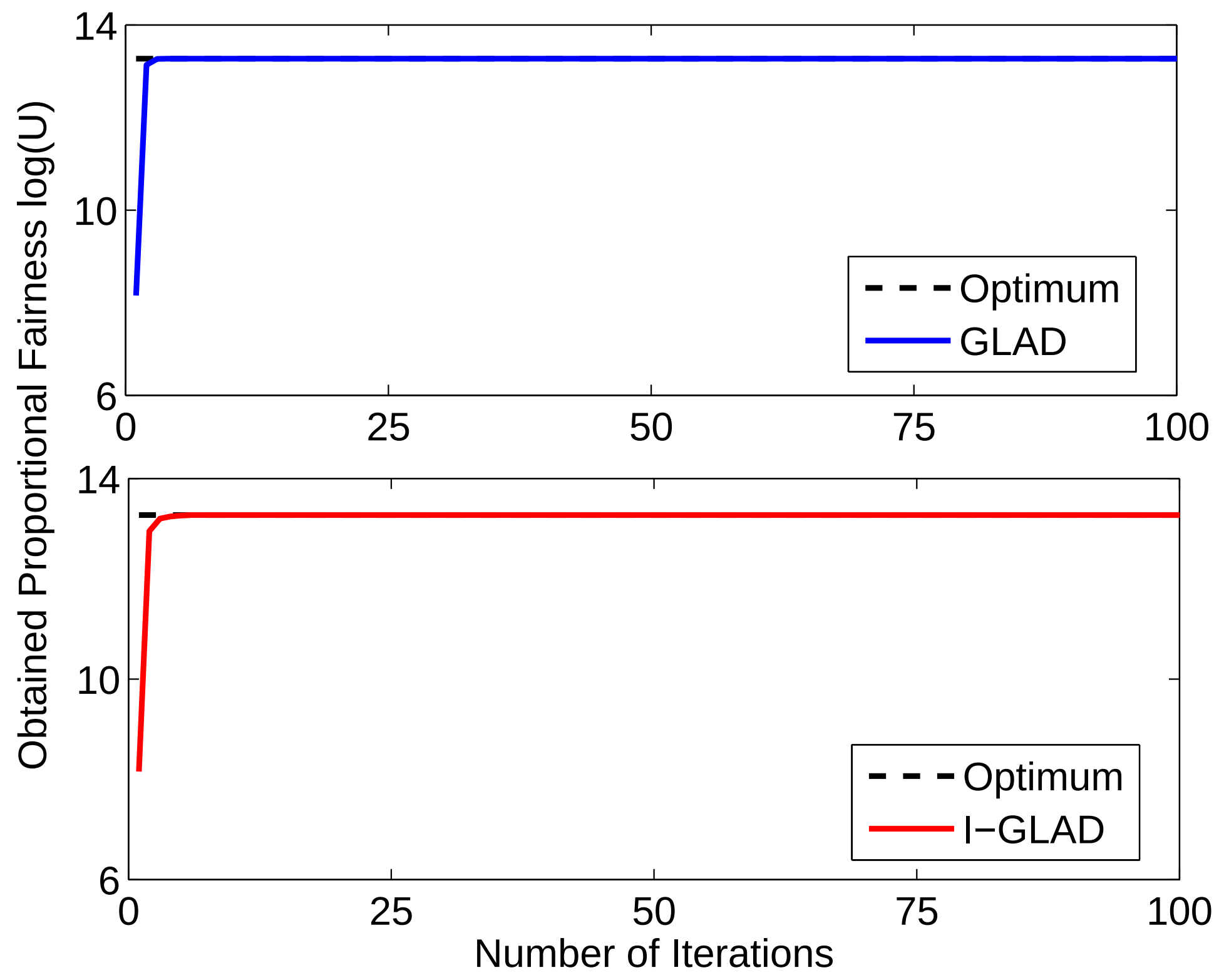

Fig. 6. Obtained proportional fairness and number of iterations for GLAD and I-GLAD with $\beta \to \infty$.

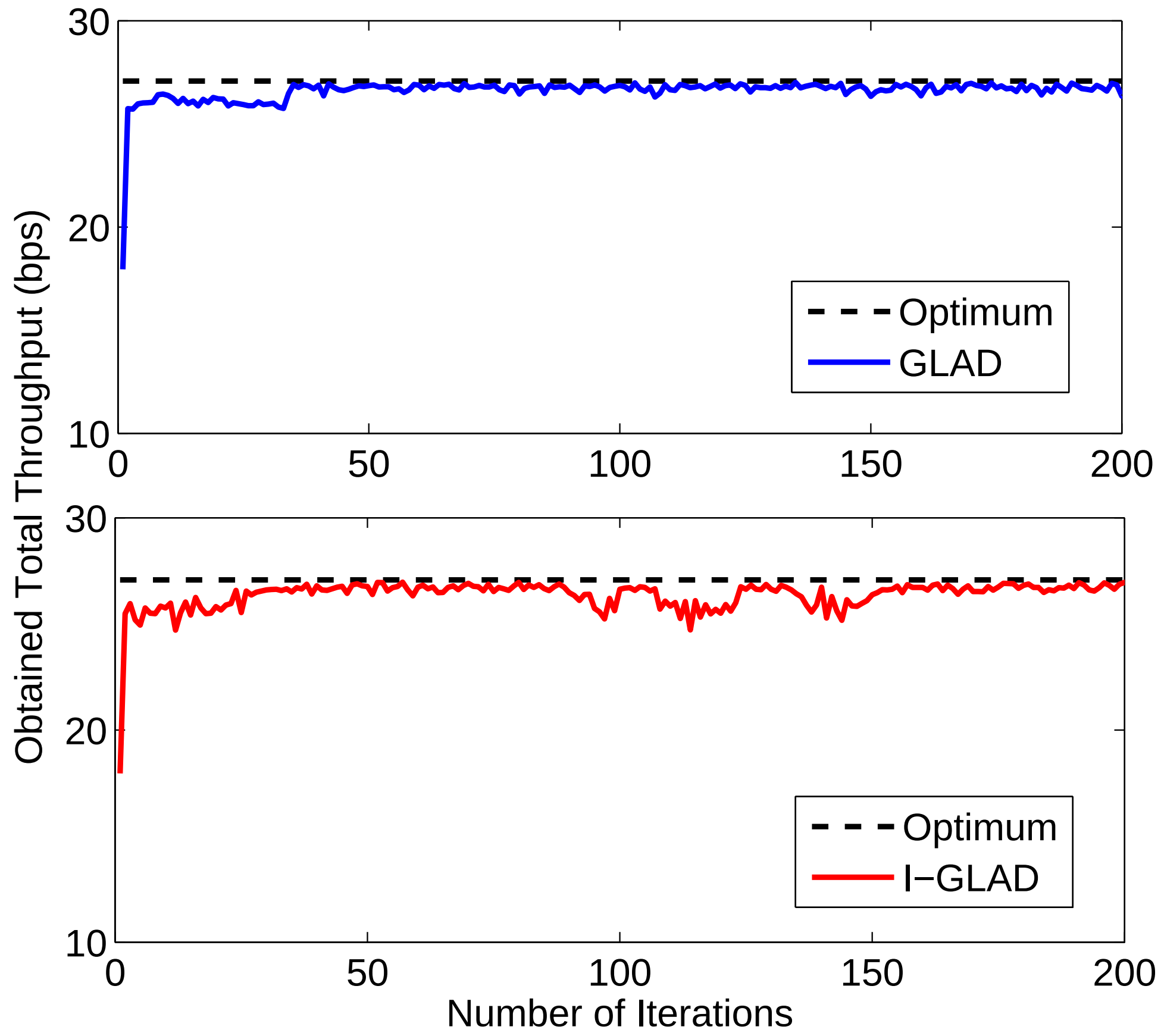

Fig. 7. Obtained total throughput and number of iterations for GLAD and I-GLAD with $\beta \to \infty$.

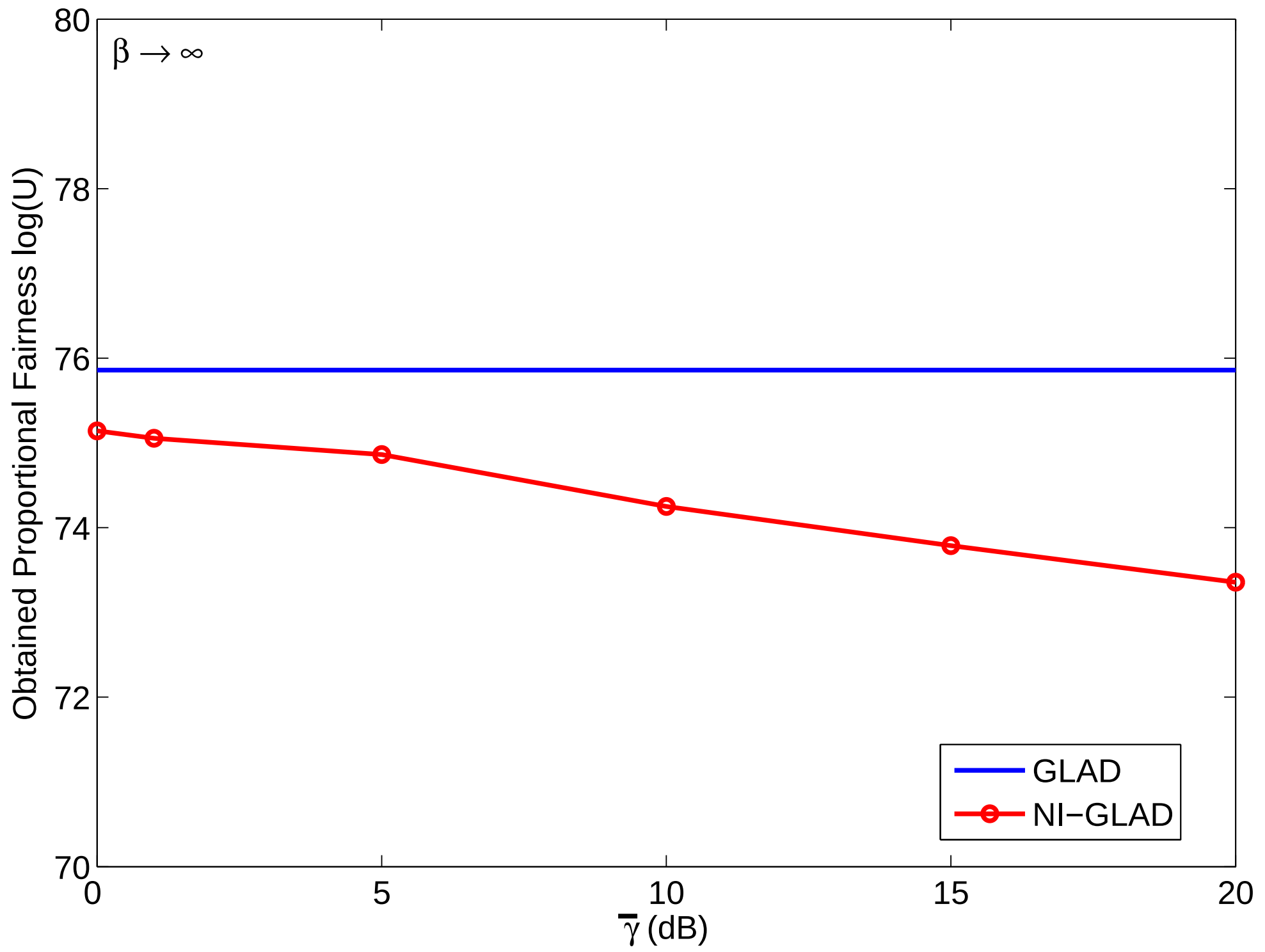

Fig. 8. Effect of $\overline{\gamma}$ on the performance of NI-GLAD for the proportional fairness utility.

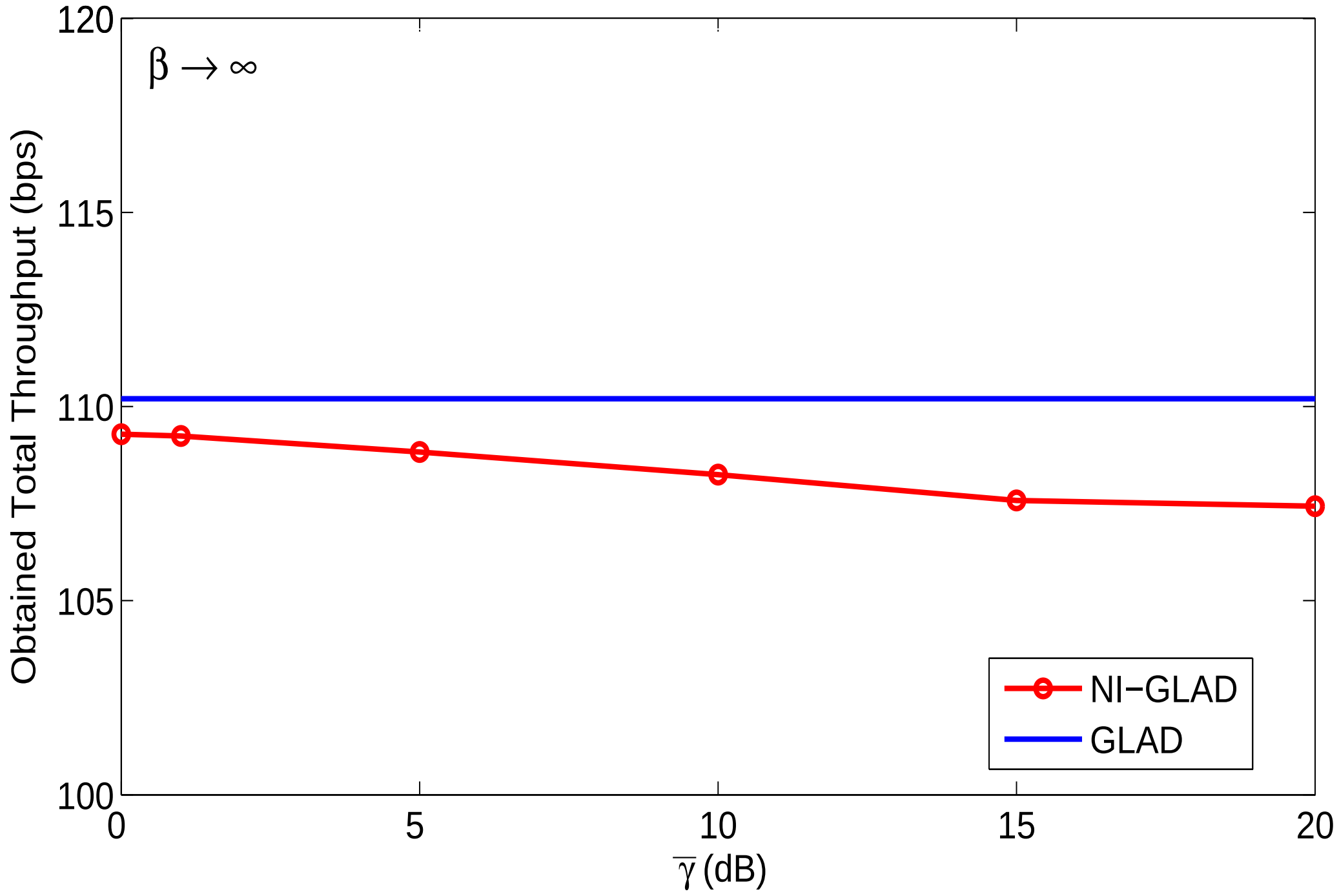

Fig. 9. Effect of $\overline{\gamma}$ on the performance of NI-GLAD for the total throughput utility.